%% file: main.tex
\documentclass[11pt]{article}

\PassOptionsToPackage{table}{xcolor}
\usepackage[preprint]{acl}
\usepackage{times}
\usepackage{latexsym}
\usepackage[T1]{fontenc}
\usepackage[utf8]{inputenc}
\usepackage{microtype}
\usepackage{graphicx}

\usepackage{amsmath,amsfonts,amssymb}
\usepackage{booktabs}
\usepackage{tabularx}
\usepackage{threeparttable}
\usepackage{siunitx}
\usepackage{algorithm}
\usepackage{algpseudocode}
\usepackage{array}
\usepackage{pifont}
\usepackage{xspace}
\usepackage{multirow}
\usepackage{newfloat}
\usepackage{placeins}
\usepackage{listings}
\usepackage{tcolorbox}
\usepackage{amssymb}
\tcbuselibrary{breakable,listings,skins}

\newcolumntype{L}{>{\raggedright\arraybackslash}X}
\newcolumntype{P}[1]{>{\raggedright\arraybackslash}p{#1}}
\newcolumntype{Z}{>{\centering\arraybackslash}p{0.105\linewidth}}
\newcolumntype{Y}{>{\raggedright\arraybackslash}X}
\newcommand{\method}{\textsc{AuditCoder}\xspace}

\title{AuditCoder: Process-Auditable Code Generation
through Responsibility-Preserving Task Graphs}
\author{
  KangjieHuang \quad Hui Zhou \quad Chen Lyu\thanks{Corresponding author.} \\
  School of Computer Science and Artificial Intelligence, \\
  Shandong Normal University \\
  \texttt{202411000534@stu.sdnu.edu.cn} \quad
  \texttt{2026217047@stu.sdnu.edu.cn} \\
  \texttt{lvchen@sdnu.edu.cn}
}

\begin{document}

\maketitle

\begin{abstract}
Plans can guide code generation without recording which code implements each subproblem, leaving reviewers to reconstruct these links during handoff and maintenance.
We present \method{}, a multi-agent system for process-auditable code generation that jointly constructs a program and its responsibility record.
A contract-annotated task graph assigns stable IDs before implementation and preserves code ownership through generation and assembly.
The same IDs connect each planning commitment to its code, versioned check results, and repair history, supporting human inspection and automated checks of record completeness and repair decisions.
When failure evidence supports a node or branch, bounded repair declares an editable scope, records any expansion, and leaves code outside the final scope unchanged.
The system abstains when no boundary is supported.
On a 200-task APPS sample and ClassEval, \method{} remains competitive with CodeSim across two models under a one-round feedback limit.
The DeepSeek-based system also remains competitive on 22 exploratory cross-domain tasks.
An audit of the DeepSeek APPS run finds that, averaged across tasks, 97.25\% of final leaf records retain nonempty code, contract, provenance, and check fields.
All 44 repair attempts follow the recorded accept/rollback policy, and 30 pass held-out tests, with scopes averaging 22\% of graph nodes. Anonymous repository:  \url{https://anonymous.4open.science/r/AuditCoder/}.
\end{abstract}

\input{sections/introduction}
\input{sections/related_work}
\input{sections/method}
\input{sections/experiment}

\input{sections/conclusion}

\input{sections/limitations}

\bibliography{references}

\clearpage
\appendix
\input{sections/appendix}

\end{document}

%% file: sections/introduction.tex
\section{Introduction}
\label{sec:introduction}

Trustworthy delivery of LLM-generated code requires developers to understand the implementations they accept and the evidence supporting them. Sonar's 2026 developer survey reports that 38\% of respondents find AI-generated code more demanding to review than code written by colleagues~\citep{sonar2026stateofcode}. During handoff and maintenance, reviewers need to connect code to its intended role, examine its checks, and understand later changes. We study \emph{process-auditable code generation}: jointly producing a program and an inspectable record linking explicit planning commitments to their implementations, check results, and repair history.

\begin{figure*}[t]
    \centering
    \includegraphics[
    width=\textwidth,
    trim=0bp 12bp 0bp 12bp,
    clip
]{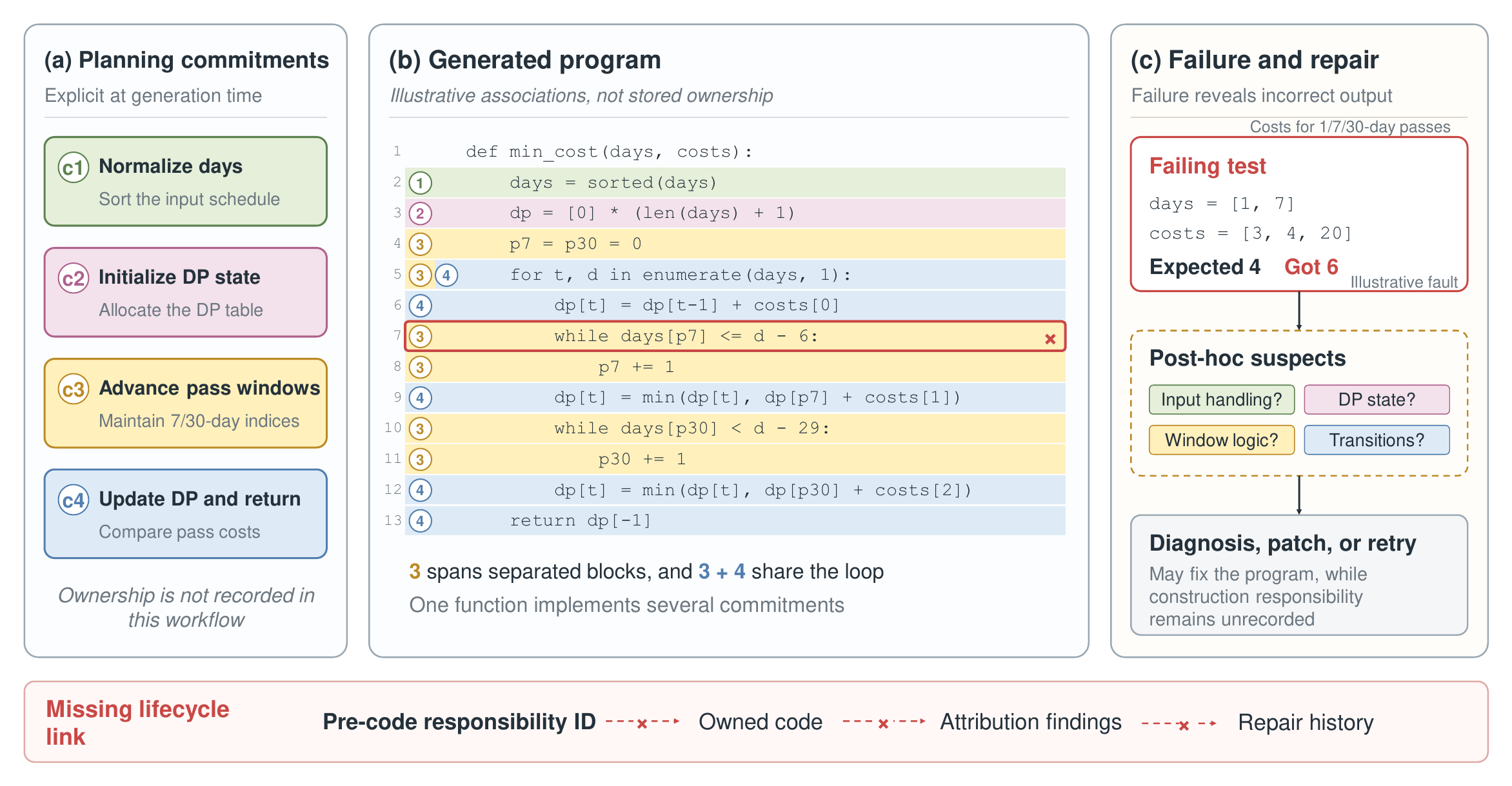}
    \caption{Decision--code responsibility misalignment. Window maintenance spans separate statements and shares a loop with DP updates. Colors indicate associations that the workflow does not record as code ownership.}
    \label{fig:decision-code-misalignment}
\end{figure*}

A plan and a finished program may not preserve these connections. One planned subproblem may span several helper functions or non-contiguous statements, while one function may implement several subproblems. In Figure~\ref{fig:decision-code-misalignment}, pass-window maintenance spans separated statements and shares a loop with dynamic-programming updates. Without recorded ownership, reviewers must reconstruct these links during code review and failure diagnosis. We call this gap \emph{decision--code responsibility misalignment}.

Existing systems use plans, execution feedback, and traceability links to guide code generation and revision~\citep{islam2024mapcoder,islam-etal-2025-codesim,hong-etal-2025-data,chen2026tracedev}. We investigate how generation-time responsibility can support human review, automated record audits, and controlled repair while retaining competitive functional performance.

The key is to construct code and its responsibility record together. A \emph{responsibility unit} identifies a planned subproblem with an explicit goal and interface before implementation. Its identity persists as code is generated, checked, and revised. Reviewers can follow these links from code to the commitments it implements and inspect the findings used to accept or revise it. Automated audits can check record completeness and consistency with the recorded repair policy. When failure evidence implicates one or more units, their owned code provides a candidate region for \emph{bounded repair}, which declares permitted changes and preserves code outside the final scope.

Maintaining these links during construction raises three challenges. First, responsibility boundaries must be chosen before code exists. Coarse units weaken localization, while overly fine units add interfaces and can split a global invariant. Second, generation and assembly introduce helper functions and shared control flow, so ownership must remain aligned with the evolving implementation and its versions. Third, failure evidence can implicate an implementation, an interface interaction, or the decomposition itself. The system must distinguish these cases before deciding which responsibilities and code to revise.

We introduce \method{}, a multi-agent generator built around a \emph{contract-annotated task graph}. The planner assigns stable IDs to subgoals with explicit contracts, dependencies, input provenance, and local check requirements before code generation. Algorithm-specialized agents implement node-owned code. An ownership map assigns every registered code unit, including helper functions and orchestration, to exactly one responsibility node and is preserved through assembly. Local verifier judgments and execution findings are attached to the corresponding IDs and code versions.

After failure, deterministic rules use current execution evidence and compatible local judgments to identify implicated nodes. Scope selection chooses a node for regeneration or a supported branch for replanning, or abstains. Integration changes require recorded scope expansions, and code outside the final scope is reused unchanged. Revalidation on the same development checks determines acceptance or rollback. Each attempt records its evidence, scope, code changes, and outcome. Figure~\ref{fig:auditcoder-lifecycle} shows the workflow (a) and how responsibility links persist through repair (b).

We evaluate functional performance with two backbone models on a fixed sample of 200 APPS tasks~\citep{hendrycks2021apps} and on ClassEval~\citep{du2024classeval}. With each method allowed at most one feedback-conditioned intervention per task, \method{} achieves 89.0--89.5\% and 90.0--92.0\% fixed-run \texttt{pass@1}, respectively, remaining competitive with CodeSim. In three exploratory suites totaling 22 tasks, the DeepSeek-based system matches CodeSim on scientific computing and aerospace/communications and solves one fewer finance task.

A read-only audit of the 200 DeepSeek APPS records finds that, averaged across tasks, 97.25\% of final leaf responsibility records retain nonempty code, contract, provenance, and check fields. Of 52 initially failing tasks, 44 receive a node-or-branch repair scope, and 30 repairs pass held-out tests. Across these 44 attempts, scopes average 22\% of graph nodes. All 44 transactions follow the accept/rollback policy. All 14 rollbacks restore the audited code and check-result snapshots. In a separate 100-task injection study, manual inspection confirms that 88 of 93 selected scopes contain the injected fault's owner.

\noindent\textbf{Contributions.}
\textbf{(1)} We formulate process-auditable code generation around decision--code responsibility misalignment.
\textbf{(2)} We introduce \method{}, which jointly constructs code and responsibility records for inspection, localization, and bounded repair.
\textbf{(3)} We evaluate functional performance, record retention, and repair behavior through benchmark comparisons, a construction-record audit, and controlled fault injection.

%% file: sections/related_work.tex
\section{Related Work}
\label{sec:related-work}

\paragraph{Planning and agent-based generation.}
Self-planning separates planning from implementation, while Parsel and FunCoder compose subfunctions~\citep{jiang2024selfplanning,zelikman2023parsel,chen2024funcoder}. AgentCoder coordinates programming and testing~\citep{huang2023agentcoder}, MapCoder uses plans to guide debugging~\citep{islam2024mapcoder}, and CodeSim verifies plans and debugs code through simulation~\citep{islam-etal-2025-codesim}. CodeTree retains strategies and feedback at candidate-program nodes~\citep{li-etal-2025-codetree}. Data Interpreter associates tasks with executable actions and results, reusing unchanged task prefixes after replanning~\citep{hong-etal-2025-data}.

\paragraph{Localized debugging and repair.}
LDB examines basic-block runtime states~\citep{zhong2024debug}, and MGDebugger decomposes faulty code into subfunctions for bottom-up debugging~\citep{shi2026mgdebugger}. Repository repair systems localize and revise existing code~\citep{yang2024sweagent,xia2025agentless}, while CodePlan propagates edits using dependency and change-impact analysis~\citep{bairi2024codeplan}. \method{} uses ownership established during generation to select a declared repair region and preserve code outside the final scope. Its repair record links the evidence, permitted changes, and outcome to the affected planning commitments.

\paragraph{Development traceability.}
TraceDev maintains requirement--design--code links during development, with test feedback handled outside its traceability graph~\citep{chen2026tracedev}. \method{} links delivered code, versioned check results, and repair transactions through planning-time responsibility IDs. Table~\ref{tab:workflow-comparison} compares working units, lifecycle links, and declared repair scopes.

\begin{table}[tb]
\centering
\begingroup
\fontsize{8.5}{10.5}\selectfont
\setlength{\tabcolsep}{1pt}
\setlength{\arrayrulewidth}{0.35pt}
\renewcommand{\arraystretch}{1.16}
\definecolor{RWHeader}{HTML}{D9D9D9}
\definecolor{RWMethod}{HTML}{F0F0F0}
\definecolor{RWRule}{HTML}{A0A0A0}
\definecolor{RWOurs}{HTML}{DDEBF7}
\definecolor{RWYes}{HTML}{286447}
\definecolor{RWMuted}{HTML}{666666}
\arrayrulecolor{RWRule}
\newcommand{\acRWUnspecified}{\textcolor{RWMuted}{---}}
\newcommand{\acRWSpecified}{\textcolor{RWYes}{$\checkmark$}}
\begin{tabularx}{\linewidth}{|>{\columncolor{RWMethod}}c|>{\centering\arraybackslash}X|c|c|}
\hline
\rowcolor{RWHeader}
\textbf{Method}
& \mbox{\textbf{Working unit}}
& \mbox{\textbf{Lifecycle link}}
& \mbox{\textbf{Declared scope}} \\
\hline
AgentCoder & Program & \acRWUnspecified & \acRWUnspecified \\ \hline
MapCoder & Plan/program & \acRWUnspecified & \acRWUnspecified \\ \hline
CodeSim & Plan/program & \acRWUnspecified & \acRWUnspecified \\ \hline
CodeTree & Candidate & \acRWUnspecified & \acRWUnspecified \\ \hline
Data Interpreter & Task/action & \acRWUnspecified & \acRWUnspecified \\ \hline
TraceDev & Artifact & \acRWUnspecified & \acRWUnspecified \\ \hline
MGDebugger & Subfunction & \acRWUnspecified & \acRWUnspecified \\ \hline
\rowcolor{RWOurs}
\textbf{\method{}} & \mbox{\textbf{Responsibility}}
& \acRWSpecified & \acRWSpecified \\
\hline
\end{tabularx}
\endgroup
\caption{Working units and specified lifecycle mechanisms.
\textbf{Lifecycle link}: a planning-time responsibility ID joining
 delivered code, versioned check results, and repair history.
\textbf{Declared scope}: a predeclared editable region, logged
 expansions, and unchanged code outside the final scope.
$\checkmark$: specified. ---: the full linkage or protocol is not
 specified in the cited workflow.}
\label{tab:workflow-comparison}
\end{table}

%% file: sections/method.tex
\providecommand{\acSem}{\mathsf{Sem}}
\providecommand{\acIfc}{\mathsf{Ifc}}
\providecommand{\acRes}{\mathsf{Res}}
\providecommand{\acCmt}{\mathsf{Cmt}}
\providecommand{\acImpl}{\mathsf{Impl}}
\providecommand{\acExec}{\mathsf{exec}}
\providecommand{\acJudg}{\mathsf{judg}}
\providecommand{\acRbar}{\overline{\mathcal R}}

\section{Method}
\label{sec:method}

\begin{figure*}[t]
    \centering
    \includegraphics[width=\textwidth]{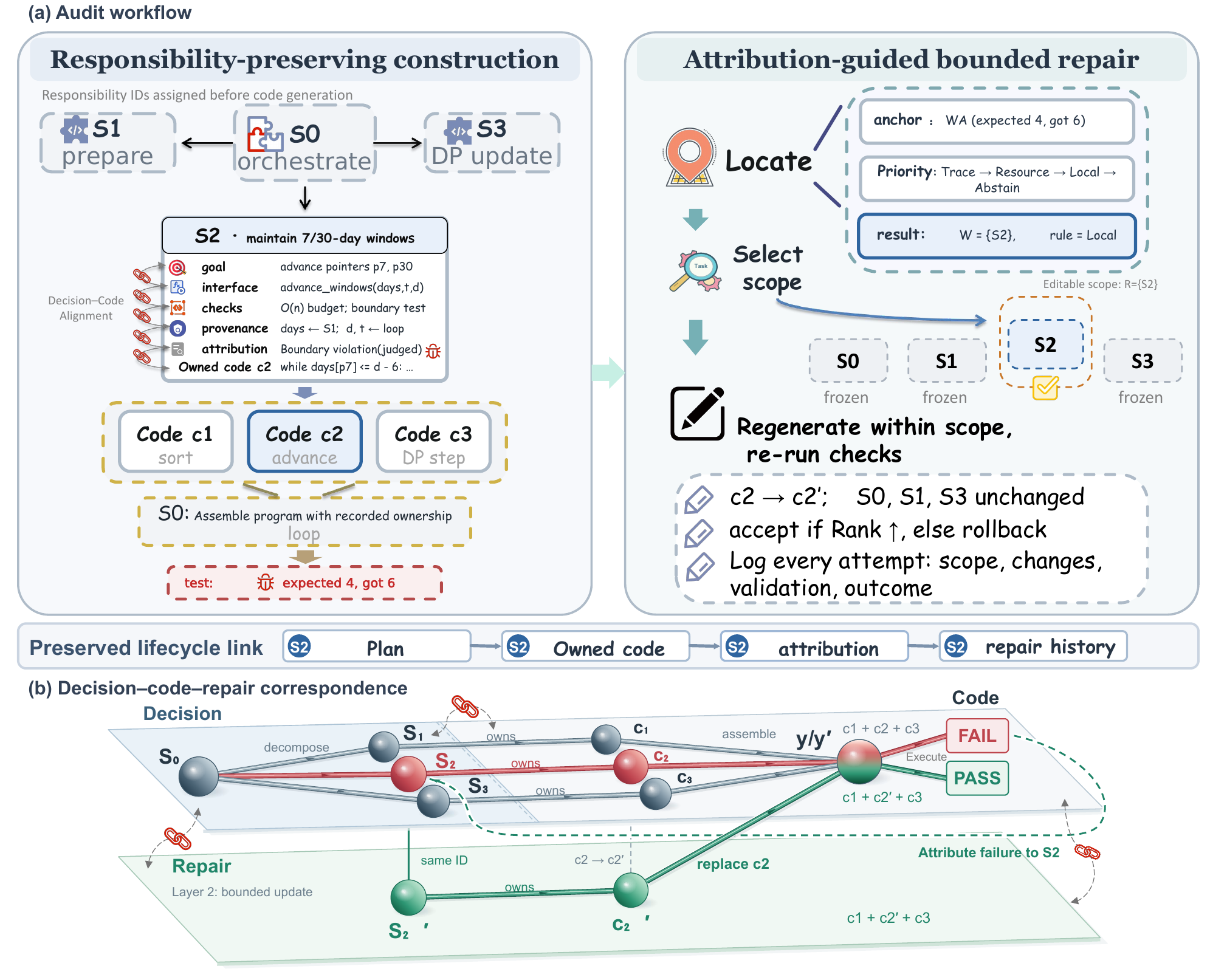}
    \caption{Responsibility-preserving generation and repair.
    (a) The workflow links planning, code generation, checks, and repair.
    (b) Replacing $c_2$ with $c'_2$ yields a candidate for revalidation
    while preserving $c_1$, $c_3$, and \texttt{S0}'s orchestration.
    \texttt{S2}$'$ retains \texttt{S2}'s responsibility ID.
    Each attempt records check results and acceptance or rollback.}
    \label{fig:auditcoder-lifecycle}
\end{figure*}

Given a natural-language task $x$, \method{} returns a program $y$ and a construction trace $\mathcal A=(\mathcal G,\mathcal S,\mathcal E,\mathcal H)$ containing the task graph, responsibility records, versioned findings, and repair history. Figure~\ref{fig:auditcoder-lifecycle} shows the workflow (a) and the responsibility links across code revisions (b). Planning, generation, assembly, verification, and replanning use LLM calls. Localization and scope selection apply deterministic rules to the resulting records. Algorithm~\ref{alg:auditcoder} gives the full workflow.

\subsection{Contract-Annotated Task Graph}
\label{subsec:contract-graph}

The planner records task I/O constraints $\mathcal P$, algorithmic assumptions and complexity targets $\mathcal B$, development checks $\mathcal Q$, and a responsibility graph
\begin{equation}
    \mathcal G=(\mathcal V,\mathcal D_h,\mathcal D_d,\mathcal T).
    \label{eq:ac-task-graph}
\end{equation}
The rooted tree $\mathcal D_h$ defines decomposition branches, with root $v_0$. The separate dependency DAG $\mathcal D_d$ records which nodes consume another node's outputs or implementation. It determines a compatible generation order $\mathcal T$ for leaf code owners. Branch selection uses the hierarchy subtree $S_h(b)$, consisting of $b$ and its descendants in $\mathcal D_h$.

Before implementation, each node receives a stable ID and records its goal $g_i$, executor category $\tau_i$, contract $F_i$, input/output slots, provenance $\rho_i$, complexity budget $C_i$, and local checks $q_i$. Contracts specify signatures, preconditions, postconditions, and invariants. Provenance identifies the provider of each input slot and induces a slot graph $\mathcal G_\rho$ for tracing recorded data flow. Code owners form $\mathcal U\subseteq\mathcal V$, including internal and root orchestration owners. Structural nodes own no code. In Figure~\ref{fig:auditcoder-lifecycle}(a), \texttt{S2} owns window maintenance and \texttt{S0} owns orchestration.

The ordered checks $\mathcal Q=(\chi_1,\ldots,\chi_n)$ contain task examples followed by planner-derived boundary cases. Their inputs, expected outputs, and order remain fixed throughout validation and repair. Held-out evaluation tests are excluded.

\subsection{Ownership-Preserving Generation}
\label{subsec:generation-assembly}

A fixed registry maps $\tau_i$ to a prompt-and-parser executor $a_i$. This records who implements a node separately from which responsibility owns its code. Following $\mathcal T$, executors generate code $c_i$ from the node's goal, contract, inputs, provenance, complexity budget, and task-level assumptions. Each implementation is locally checked before assembly. The assembler emits internal and root orchestration code and preserves the ownership map
\begin{equation}
    \mu:\mathsf{Unit}_{\mathrm{reg}}\longrightarrow\mathcal U.
    \label{eq:ac-ownership}
\end{equation}
Every registered executable unit, including helper functions and orchestration statements, has exactly one owner. Assembly preserves these ownership assignments and provider bindings, producing $y=\Omega(\mathcal G,\{(v_i,c_i)\}_{v_i\in\mathcal U})$ (Figure~\ref{fig:auditcoder-lifecycle}(b)).

\paragraph{Versioned findings.}
Local verification checks contracts, local requirements, and budgets. End-to-end execution on $\mathcal Q$ produces global results $r_g$. Both append findings to $\mathcal E$, recording owners, code version, verdict, checked obligation, witness, and evidence kind. Executed findings ($\acExec$) carry runtime witnesses. Judged findings ($\acJudg$) contain verifier assertions without execution witnesses. In the reported runs, node-local checks are judged, while executed evidence comes from end-to-end checks, tracebacks, and resource measurements. Node-indexed findings $r_i$ collect records associated with $v_i$. Superseded findings remain in the trace, but only current findings enter localization.

\subsection{Failure-to-Responsibility Localization}
\label{subsec:failure-localization}

\paragraph{Anchor and rule priority.}
The first failing check $\chi^\star$ in $\mathcal Q$ anchors localization. The locator abstains ($z=\bot$) or returns
\begin{equation}
    z=(W,\ell,\kappa,\widehat{\mathcal E}),
    \label{eq:ac-localization}
\end{equation}
where $W\subseteq\mathcal V$ is a nonempty set of implicated responsibilities, $\ell$ identifies the rule, $\kappa$ grades the recorded evidence, and $\widehat{\mathcal E}$ contains supporting findings. Rules follow the priority \textsc{Trace} $\rightarrow$ \textsc{Resource} $\rightarrow$ \textsc{Local}. \textsc{Trace} maps a generated traceback frame through $\mu$. \textsc{Resource} requires an owner-attributed budget violation. \textsc{Local} retains local findings compatible with the failing case.

\paragraph{Compatibility and coherence.}
A compatible local finding must be current, rejecting, and tied to a specific obligation. Wrong answers and runtime errors admit semantic, interface, or planning-commitment findings. Time and memory limits admit resource or planning-commitment findings. The constrained slot must reach the failing observable in $\mathcal G_\rho$, and the anchor input must satisfy the finding's witness condition. If a finding's input violates its owner's precondition, attribution shifts to the provider--consumer interface. A finding contradicted by current execution evidence is excluded. Among compatible local findings, the rule prefers those with observed execution applicability and otherwise uses judged records.

All commitment findings must name the same planning decision, whose subtree contains $W$. Otherwise, $W$ must be a singleton or form a connected interface provider-pair graph. Unsupported or competing explanations lead to abstention. The defect class $\delta$ is commitment ($\acCmt$) when a planning decision is implicated, interface ($\acIfc$) when an interface is implicated, and implementation ($\acImpl$) otherwise. Grades are ordinal (high, medium, or low). Appendix~\ref{app:compat} gives the full rules.

In Figure~\ref{fig:auditcoder-lifecycle}(a), a seven-day pass should cover days 1 and 7. A compatible verifier judgment rejects \texttt{S2}'s premature removal of day 1, so \textsc{Local} returns $W=\{\texttt{S2}\}$.

\subsection{Attribution-Guided Bounded Repair}
\label{subsec:bounded-repair}

\paragraph{Scope selection.}
For an implementation defect, the selector first tries an eligible singleton. For interface or commitment defects, it searches for a supported branch. Non-root ancestors covering $W$ are examined from deepest to shallowest:
\begin{equation}
    \mathcal R_0=
    \begin{cases}
        \{v\}, & \text{eligible singleton},\\
        S_h(b^\star), & \text{deepest eligible ancestor},\\
        \bot, & \text{no supported boundary}.
    \end{cases}
    \label{eq:ac-initial-scope}
\end{equation}
Eligibility depends on the defect class and supporting obligations (Algorithm~\ref{alg:scope}). Dependencies crossing the boundary identify integration points. A scope expansion must record the affected owners and supporting findings before their code is changed. Unsupported expansion or a scope leaving no frozen code owner causes abstention.

\paragraph{Regeneration and acceptance.}
Let $\mathcal R$ be the final declared scope and $\acRbar=\mathcal V\setminus\mathcal R$ its complement. The system regenerates a selected node or replans and regenerates a selected branch. It assembles the replacement with the preserved code subject to
\begin{equation}
    c'_v=c_v\qquad\forall v\in\acRbar\cap\mathcal U.
    \label{eq:ac-frozen-complement}
\end{equation}
Surviving responsibilities retain their IDs, and new responsibilities receive fresh IDs. Ownership, dependency order, interfaces, and provenance are rebuilt and checked. Revised outputs can affect dependent nodes even when their code is frozen. All development checks are therefore rerun in the same order. Acceptance requires strict lexicographic improvement in
\begin{equation}
    \begin{aligned}
        \mathrm{Rank}(r)=\bigl(&\mathbf 1[\text{all checks pass}],\\
        &n_{\mathrm{pass}}^{\mathrm{ext}},n_{\mathrm{pass}}^{\mathrm{int}},
         n_{\mathrm{clean}}^{\mathrm{ext}},n_{\mathrm{clean}}^{\mathrm{int}}\bigr).
    \end{aligned}
    \label{eq:ac-acceptance-rank}
\end{equation}
Here $n_{\mathrm{pass}}$ and $n_{\mathrm{clean}}$ count passed checks and clean exits, respectively. Superscripts distinguish task examples (ext) from planner-derived checks (int). Otherwise, the active program and required snapshots are restored. In Figure~\ref{fig:auditcoder-lifecycle}(b), replacing $c_2$ with $c'_2$ within $\mathcal R=\{\texttt{S2}\}$ produces the reassembled candidate $y'$ for revalidation while preserving $c_1$, $c_3$, and \texttt{S0}'s orchestration.

\paragraph{Repair history.}
Each attempted repair appends its localization result, initial and final scope, integration exceptions, graph and code changes, revalidation results, decision, and rollback status to $\mathcal H$, linked to the affected responsibilities. When localization or scope selection abstains, no repair call is made and the active program remains unchanged. Appendix~\ref{app:schema} specifies record schemas, and Appendix~\ref{app:repair-locality} details snapshot restoration.

%% file: sections/experiment.tex

\section{Experiments}
\label{sec:experiments}

\definecolor{ExpHeader}{HTML}{D9D9D9}
\definecolor{ExpSubHeader}{HTML}{F2F2F2}
\definecolor{ExpLine}{HTML}{7F7F7F}
\definecolor{ExpBase}{HTML}{FFF2CC}
\definecolor{ExpGraph}{HTML}{DDEBF7}
\definecolor{ExpAdaptive}{HTML}{E2F0D9}
\definecolor{ExpAudit}{HTML}{FCE4D6}
\definecolor{ExpAux}{HTML}{F4F6F8}
\definecolor{ExpBest}{HTML}{FBE5A2}
\arrayrulecolor{ExpLine}
\newcommand{\best}[1]{\cellcolor{ExpBest}\textbf{#1}}
\newcommand{\expsectionrow}[2]{%
  \rowcolor{#1}\multicolumn{#2}{|l|}{\textbf{\strut}}\\[-1.55ex]
}
\newcommand{\expblock}[3]{%
  \rowcolor{#1}\multicolumn{#2}{|l|}{\textbf{#3}}\\ \hline
}

Our evaluation addresses three questions. \textbf{RQ1} examines functional performance, component contributions, and intervention budgets. \textbf{RQ2} examines record completeness, fault-owner containment, and repair transactions. \textbf{RQ3} examines applicability and transfer.

\subsection{Setup}
\label{subsec:exp-setup}

\paragraph{Tasks and methods.}
We use a fixed APPS sample of 200 Python tasks: 50 introductory, 100 interview, and 50 competition problems~\citep{hendrycks2021apps}. ClassEval adds class-level generation with interdependent methods and shared state, with 100 records per model--method run~\citep{du2024classeval}. Baselines are AgentCoder~\citep{huang2023agentcoder}, MapCoder~\citep{islam2024mapcoder}, and CodeSim~\citep{islam-etal-2025-codesim}. Table~\ref{tab:functional-results} also reports graph variants with task planning, algorithm-specialized agents, and bounded repair. We use \texttt{deepseek-v4-flash} (DS) and Qwen3.6-plus (QW)~\citep{alibaba2026qwen36plus}, holding task inputs, decoding policy, execution harness, and resource limits fixed across methods. Appendix~\ref{app:exp-details} gives settings and exclusions. Reproduction commands accompany the Code and Data Supplement.

\paragraph{Intervention and scoring.}
In the primary setting, each complete method is allowed at most one feedback-conditioned debugging, retry, or repair round ($K{=}1$). Variants without repair receive none, and an abstention makes no repair call. We use $K{=}1$ to compare performance after the first feedback-driven revision and audit each attempted repair against its pre-repair state. This aligns feedback opportunities, not token use or model calls. Fixed-run \texttt{pass@1} scores the single output retained after the method's sequence. In generation runs, hidden tests are used only for final evaluation.

A separate APPS setting removes the round cap and stops after four consecutive rounds without an increase in the number of passed tests. An increase resets the counter. Each method retains its own candidate-update rule. The process audit uses only the 200 DS APPS traces from the primary \method{} run. A separate fault-injection experiment evaluates one-round repair from selected, initially passing programs. The exploratory transfer checks are described under RQ3.

\subsection{RQ1: Functional Performance}
\label{subsec:main-results}

\begin{table}[tb]
    \centering
    \small
    \setlength{\tabcolsep}{4pt}
    \renewcommand{\arraystretch}{1.12}
    \begin{tabular}{|>{\raggedright\arraybackslash}p{0.41\columnwidth}|>{\centering\arraybackslash}p{0.09\columnwidth}|>{\centering\arraybackslash}p{0.09\columnwidth}|>{\centering\arraybackslash}p{0.09\columnwidth}|>{\centering\arraybackslash}p{0.09\columnwidth}|}
        \hline
        \rowcolor{ExpHeader}
        \textbf{Method} & \multicolumn{2}{c|}{\textbf{APPS}} & \multicolumn{2}{c|}{\textbf{ClassEval}} \\
        \cline{2-5}
        \rowcolor{ExpSubHeader}
         & \textbf{DS} & \textbf{QW} & \textbf{DS} & \textbf{QW} \\
        \hline
        \rowcolor{ExpBase}
        \multicolumn{5}{|l|}{\textbf{Multi-agent baselines}} \\
        AgentCoder & 65.0 & 70.0 & 80.0 & 82.0 \\
        \rowcolor{ExpAux}
        MapCoder   & 76.5 & 80.0 & 89.0 & 92.0 \\
        CodeSim    & 87.0 & 88.5 & \best{91.0} & \best{95.0} \\
        \hline
        \rowcolor{ExpGraph}
        \multicolumn{5}{|l|}{\textbf{Graph-based variants}} \\
        Task plan & 39.5 & 62.5 & 58.0 & 62.0 \\
        \rowcolor{ExpAux}
        \quad + algorithm agents & 74.0 & 70.0 & 71.0 & 71.0 \\
        \rowcolor{ExpGraph}
        \method{} (+ repair) & \best{89.0} & \best{89.5} & 90.0 & 92.0 \\
        \hline
        \rowcolor{ExpAdaptive}
        \multicolumn{5}{|l|}{\textbf{Adaptive multi-round repair (APPS)}} \\
        AgentCoder & 88.5 & 92.0 & -- & -- \\
        \rowcolor{ExpAux}
        MapCoder   & 91.5 & 96.0 & -- & -- \\
        CodeSim    & 93.0 & 97.0 & -- & -- \\
        \rowcolor{ExpAdaptive}
        \method{} & \best{94.5} & \best{97.5} & -- & -- \\
        \hline
    \end{tabular}
    \caption{Fixed-run \texttt{pass@1} (\%) on APPS (200 tasks) and ClassEval (100 records). The first two blocks allow at most one feedback round (none for no-repair variants). The last uses adaptive multi-round repair. Shaded cells indicate section structure; highlighted cells mark the best result within each setting. --: unreported.}
    \label{tab:functional-results}
    \label{tab:apps-results}
    \label{tab:classeval-results}
\end{table}

\paragraph{Complete-system performance.}
Under the single-intervention protocol, \method{} reaches 89.0/89.5\% on APPS and 90.0/92.0\% on ClassEval with DS/QW, respectively (Table~\ref{tab:functional-results}). Its APPS advantage over CodeSim is 2.0/1.0 percentage points, or four/two tasks. On ClassEval, it trails CodeSim by 1.0/3.0 points and matches or exceeds the other baselines. The complete pipeline remains competitive on both benchmarks.

\paragraph{Component contributions.}
On APPS, adding algorithm-specialized agents raises the task-plan variant from 39.5/62.5\% to 74.0/70.0\%. Bounded repair further raises performance to 89.0/89.5\%. The repair increment includes an additional feedback opportunity, so it does not isolate responsibility binding from repair itself.

\paragraph{Sensitivity to repeated intervention.}
With adaptive multi-round repair, \method{} reaches 94.5/97.5\%, compared with CodeSim's 93.0/97.0\%. Its gains over the primary setting are 5.5/8.0 points. AgentCoder and MapCoder gain more, narrowing their gaps to \method{}. Functional competitiveness extends beyond one round, although the intervention budget affects the comparison.

\begin{figure}[tb]
    \centering
    \includegraphics[width=\columnwidth]{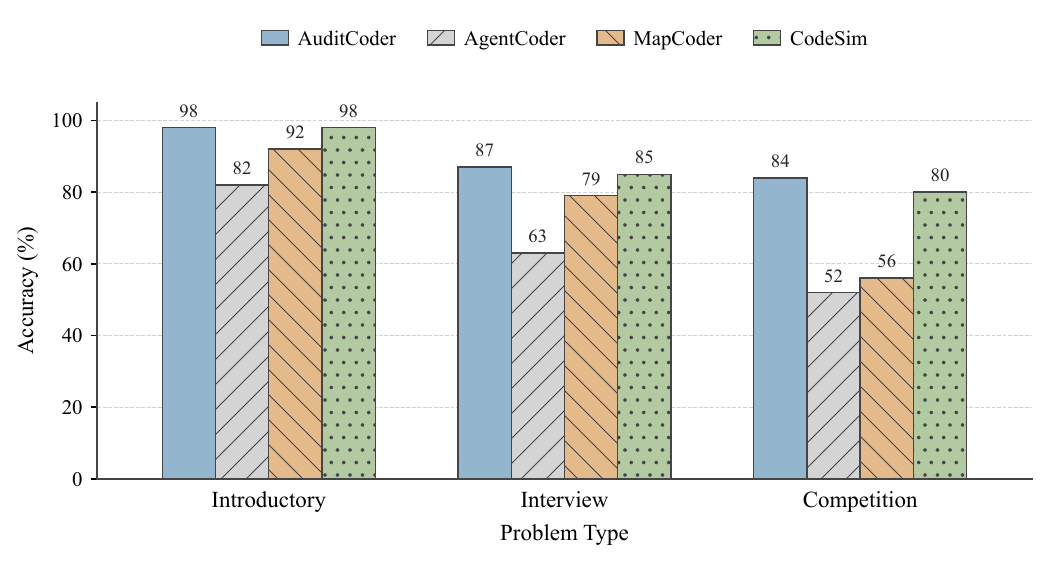}
    \caption{APPS fixed-run \texttt{pass@1} (\%) by difficulty with DS and at most one feedback round. The three groups contain 50, 100, and 50 tasks.}
    \label{fig:difficulty-accuracy}
\end{figure}

\begin{figure}[tb]
    \centering
    \includegraphics[width=\columnwidth]{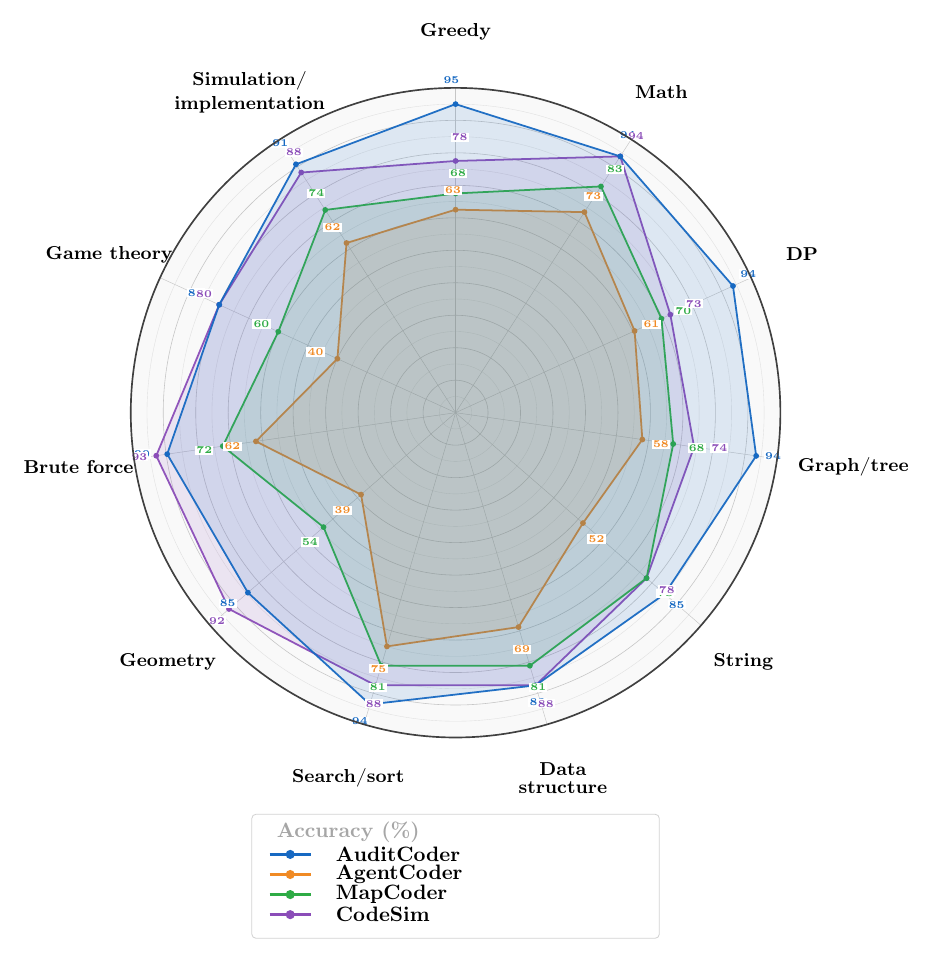}
    \caption{APPS accuracy by problem type (\%). Rings are spaced by 5 percentage points, and point labels are rounded to the nearest percentage point.}
    \label{fig:category-accuracy}
\end{figure}

\paragraph{Difficulty and problem type.}
Figure~\ref{fig:difficulty-accuracy} shows DS APPS performance by difficulty. \method{} ties CodeSim at 98\% on introductory tasks and reaches 87\% versus 85\% on interview tasks and 84\% versus 80\% on competition tasks. Each of the latter two margins represents two tasks. Across the 11 problem types in Figure~\ref{fig:category-accuracy}, \method{} leads in six categories and ties CodeSim in three. Its two deficits are geometry (84.6\% versus 92.3\%) and brute force (89.7\% versus 93.1\%).

\subsection{RQ2: Auditability and Bounded Repair}
\label{subsec:audit-metrics}

\begin{table}[tb]
\centering
\begingroup
\small
\setlength{\tabcolsep}{1.6pt}
\setlength{\arrayrulewidth}{0.35pt}
\renewcommand{\arraystretch}{1.10}
\renewcommand{\tabularxcolumn}[1]{m{#1}}
\definecolor{ATHeader}{HTML}{D9D9D9}
\definecolor{ATRule}{HTML}{A0A0A0}
\definecolor{ATTrace}{HTML}{FCE4D6}
\definecolor{ATRepair}{HTML}{DDEBF7}
\definecolor{ATStripe}{HTML}{F4F6F8}
\definecolor{ATEmphasis}{HTML}{FBE5A2}
\arrayrulecolor{ATRule}
\newcommand{\acTableThreeHighlight}[1]{\cellcolor{ATEmphasis}\textbf{#1}}
\begin{tabularx}{\linewidth}{|>{\raggedright\arraybackslash}X|c|c|c|}
\hline
\rowcolor{ATHeader}
\textbf{Metric} & $n$ & \textbf{Mean / rate} & \textbf{Med.} \\
\hline
\rowcolor{ATTrace}
\multicolumn{4}{|l|}{\textbf{Trace and graph structure}} \\
DCTC-leaf, task macro~$\uparrow$ & 200
& \mbox{0.9625 $\rightarrow$ 0.9725}
& \mbox{1.0 $\rightarrow$ 1.0} \\
\rowcolor{ATStripe}
Graph integrity~$\uparrow$ & 200 & 185/200 (.925) & -- \\
Contract--signature consistency~$\uparrow$ & 490 & 480/490 (.980) & -- \\
\hline
\rowcolor{ATRepair}
\multicolumn{4}{|l|}{\textbf{Repair attribution and governance}} \\
Localization coverage~$\uparrow$ & 52 & 44/52 (.846) & -- \\
\rowcolor{ATStripe}
Localized success~$\uparrow$ & 44 & 30/44 (.682) & -- \\
RRS (nodes) & 44 & .221 & .200 \\
\rowcolor{ATStripe}
CCR (LOC) & 44 & .380 & .392 \\
Test regressions~$\downarrow$ & 131 & \acTableThreeHighlight{0/131 (.000)} & -- \\
\rowcolor{ATStripe}
Decision compliance~$\uparrow$ & 44 & \acTableThreeHighlight{44/44 (1.000)} & -- \\
Rollback integrity~$\uparrow$ & 14 & \acTableThreeHighlight{14/14 (1.000)} & -- \\
\hline
\end{tabularx}
\endgroup
\caption{Audit of 200 DS APPS traces under the single-intervention protocol.
$n$ is each metric's denominator. DCTC values are shown before $\rightarrow$
after repair. RRS and CCR measure scope size and retained code changes.
Yellow highlights the final three audit rates.}
\label{tab:audit-metrics}
\end{table}

\paragraph{Inspectable construction records.}
Table~\ref{tab:audit-metrics} measures record structure, intervention scope, and transaction behavior. Decision--Code Trace Coverage (DCTC) is the fraction of leaf records with nonempty owned code, contract, provenance, and validation fields, averaged across tasks. It rises from 0.9625 after assembly to 0.9725 after repair. This is field completeness, not semantic correctness. Graph integrity holds for 185/200 records, requiring resolvable references, an acyclic parent relation, and an order listing each recorded leaf once. Contract--signature consistency holds for 480/490 checked instances, covering parseability, positional parameters, and uniqueness of any declared function name. The retained artifacts therefore support automated checks that expose incomplete or inconsistent construction records.

\paragraph{Scoped intervention on natural failures.}
Of 52 initially failing tasks, 44 receive a recorded node-or-branch scope and 30 repairs pass hidden tests. The other eight tasks receive no repair call. The final 178/200 passing tasks reconcile with the 89.0\% result in Table~\ref{tab:functional-results}. Across the 44 transactions, mean repair-region size (RRS) is 0.221 of graph nodes, while mean changed-code ratio (CCR) is 0.380 of final lines of code. Excluding 14 zero-delta rollbacks, accepted repairs change about 0.56 of final LOC on average. RRS describes the declared boundary, whereas CCR describes the retained textual change. A small node scope can still contain much of the program.

\paragraph{Checking repair decisions.}
All 44 transactions follow the recorded strict-improvement ordering over full pass, external/internal passes, and clean exits: 30 are accepted and 14 rolled back. All rollbacks restore the three audited snapshots: assembled code, node snippets, and development-check results. No regression appears among the 131 previously passing hidden-test observations tracked by the audit. These records allow an auditor to inspect the decision and restoration outcome separately from the program's final score. Metric definitions and snapshot checks appear in Appendices~\ref{app:audit-metric-definitions} and~\ref{app:repair-locality}.

For example, APPS task 3103 links a wrong answer and compatible local rejections to branch \texttt{S1}. Repair preserves its external \texttt{raw\_input}--to--\texttt{result} contract and freezes outer orchestration. All three task examples pass after reassembly. The record connects the selected responsibility, supporting findings, revision, and outcome (Appendix~\ref{app:case-study}).

\begin{table}[tb]
    \centering
    \begingroup
    \small
    \setlength{\tabcolsep}{2pt}
    \setlength{\arrayrulewidth}{0.4pt}
    \renewcommand{\arraystretch}{1.10}
    \arrayrulecolor{ExpLine}
    \newcommand{\tFourBest}[1]{\cellcolor{ExpBest}\textbf{#1}}

    \begin{tabularx}{\linewidth}{|>{\raggedright\arraybackslash}X|*{5}{c|}r|r|}
        \hline
        \rowcolor{ExpHeader}
        \textbf{Method}
        & \multicolumn{5}{c|}{\textbf{Repaired faults}}
        & \multicolumn{2}{c|}{\textbf{Tokens}} \\
        \cline{2-8}
        \rowcolor{ExpSubHeader}
        & \textbf{RE} & \textbf{WA} & \textbf{IC}
        & \textbf{TO} & \textbf{All}
        & \textbf{/task} & \textbf{/fix} \\
        \hline
        \rowcolor{ExpBase}
        AgentCoder$^\dagger$
        & 19 & 16 & 18 & 17 & 70 & 3{,}410 & 4{,}871 \\
        \rowcolor{ExpAux}
        MapCoder
        & \tFourBest{23} & 20 & \tFourBest{23} & 20 & 86
        & \tFourBest{2{,}026} & \tFourBest{2{,}356} \\
        CodeSim
        & \tFourBest{23} & \tFourBest{21} & 22
        & \tFourBest{21} & \tFourBest{87} & 2{,}125 & 2{,}442 \\
        \rowcolor{ExpGraph}
        \method{}
        & \tFourBest{23} & 20 & \tFourBest{23} & 19 & 85
        & 2{,}289 & 2{,}693 \\
        \hline
        \rowcolor{ExpAudit}
        \multicolumn{8}{|>{\raggedright\arraybackslash}p{\dimexpr\linewidth-2\tabcolsep-2\arrayrulewidth\relax}|}{%
            \textbf{Human-verified scope selection (\method{})}} \\
        \multicolumn{5}{|l|}{Non-abstaining / abstaining tasks}
        & \multicolumn{3}{c|}{93 / 7} \\
        \rowcolor{ExpAux}
        \multicolumn{5}{|l|}{Fault owner in selected scope}
        & \multicolumn{3}{c|}{88/93 (94.6\%)} \\
        \hline
    \end{tabularx}
    \endgroup
    \caption{One-round repair on 100 APPS interview tasks with DS (25 faults per class). RE/WA/IC/TO denote runtime exceptions, wrong answers, interface-contract violations, and timeouts. Repair-round tokens exclude post-injection local checks. /task averages over all tasks. /fix divides total tokens by successful repairs. Highlighted cells mark the best value in each column. $^\dagger$AgentCoder counts only its programmer call.}
    \label{tab:injection-repair}
\end{table}

\paragraph{Controlled fault injection.}
Natural failures arise from different initial programs and do not isolate repair behavior. Using DS and the 100 APPS interview tasks, we sample programs with each method's own pipeline and select an initially hidden-test-passing program per task and method. This selection constructs the repair cohort and is separate from fixed-run generation. We inject one fault per program, with 25 tasks in each of four classes (Table~\ref{tab:injection-repair}). Across methods, each task receives the same fault class and semantic behavior change, implemented in the corresponding code. Injection markers are bookkeeping only and hidden from all methods. Local checks are rerun after injection to obtain current findings. Each method then receives one repair/debugging round before hidden-test evaluation. For non-abstaining \method{} tasks, manual inspection checks whether the selected scope contains the responsibility unit owning the injected fault.

\paragraph{Fault-owner containment.}
Of 100 injected cases, \method{} selects scopes for 93 and abstains on seven. Manual inspection places the true fault owner within 88 of the 93 selected scopes (94.6\%), or 88.0\% of all tasks. Five selected scopes miss the fault owner. This checks the content of a localization decision beyond the coverage statistic reported for natural failures. Containment checks whether the authorized region includes the faulty owner. Repair success separately measures whether the replacement passes the tests.

\paragraph{Repair outcomes and cost.}
\method{} repairs 85 faults, versus 86 for MapCoder, 87 for CodeSim, and 70 for AgentCoder. Eight attempted repairs remain failing, alongside seven abstentions. \method{} ties for the most repairs on runtime exceptions and interface-contract violations (23/25 each), while timeouts have the fewest repairs (19/25). These counts distinguish failure to select a scope from failure to obtain a passing replacement.

Repair-round cost is 2,289 tokens per task, 7.7\% above CodeSim and 13.0\% above MapCoder, and 2,693 tokens per repaired fault. Per-task cost averages over all 100 tasks, including abstentions, while per-fix cost includes tokens spent on unsuccessful attempts. The token counts exclude post-injection local checks. Both prompt length (1,672 versus 1,566--1,572 tokens per task) and completion length (617 versus 454--559) contribute to the difference.

The two cohorts provide complementary evidence: natural runs test record completeness and transaction behavior, while injected faults test whether the selected scope contains the fault owner.

\subsection{RQ3: Applicability and Transfer}
\label{subsec:suitability-analysis}
\label{subsec:scope-cost}

\paragraph{Responsibility boundaries.}
Figure~\ref{fig:suitability-cases-app} examines contrasting cases with 10 attempts per case (Appendix~\ref{app:failure-analysis}). Stable intermediate artifacts provide useful responsibility boundaries. When correctness depends on a tightly coupled global invariant, a coarser unit, graph revision, or abstention may be appropriate. These cases illustrate boundary selection without estimating its prevalence.

\paragraph{Exploratory transfer.}
We also test 22 tasks spanning file-based finance, blind scientific computing, and source-grounded aerospace/communications, using DS, one retained candidate, and frozen hidden tests (Appendix~\ref{app:cross-domain-suites}; Figure~\ref{fig:cross-domain-results}). On QF-Small6, SciCode Blind-Small-8, and AeroComm-Small8, respectively, \method{} resolves 5/6, 7/8, and 8/8 tasks, compared with AgentCoder's 3/6, 6/8, and 5/8; MapCoder's 5/6, 7/8, and 7/8; and CodeSim's 6/6, 7/8, and 8/8. These small suites provide exploratory evidence of functional competitiveness across task forms.

%% file: sections/conclusion.tex
\section{Conclusion}
\label{sec:conclusion}

\method{} jointly constructs code and responsibility records for inspection and bounded repair, preserving code outside the declared scope. Experiments show competitive functional performance, fault-owner containment, and repair decisions that can be checked against recorded evidence. These records support code handoff and maintenance.

%% file: sections/limitations.tex
\section*{Limitations}

\paragraph{Decomposition and local evidence.}
\method{} depends on a decomposition with meaningful responsibility boundaries. Tightly coupled global invariants may require a coarser unit or graph revision. In the reported runs, node-local contract checks rely on LLM judgments, while end-to-end development checks supply execution evidence. Verifier errors can misdirect localization. Insufficient evidence leads to abstention, and selecting a scope does not ensure a successful repair.

\paragraph{Scope of the process audit.}
Table~\ref{tab:audit-metrics} audits one \texttt{deepseek-v4-flash} APPS run. Its rates have not been replicated across runs, models, or benchmarks. DCTC measures field completeness, not the correctness of contracts or attributions. Decision compliance compares accept/rollback outcomes with recorded ranks. Rollback integrity compares snapshots of assembled code, node snippets, and development-check results. These checks do not establish full intermediate-state restoration, per-node reverification of frozen code, or a complete event history.

\paragraph{Comparison protocol.}
Primary comparisons align feedback opportunities, not token use or model calls. Additional multi-round APPS results extend the accuracy comparison, while the process audit remains limited to a single-intervention run. Component ablations that add repair also add feedback, so they do not isolate the effect of responsibility binding. The small accuracy differences from CodeSim have not been tested for stability across repeated runs.

\paragraph{Task scope and human use.}
Evaluation covers algorithmic and class-level Python tasks and three small exploratory suites. Transfer to repositories with cross-file dependencies and build constraints, or to other programming languages, remains unevaluated. Manual inspection in the injection study verifies fault-owner containment. Effects of the records on developers' review accuracy, maintenance effort, and trust have not been measured.

\paragraph{Cost and reproducibility.}
In the injection study, \method{} uses 8--13\% more repair-round tokens than MapCoder and CodeSim, excluding post-injection local checks. This comparison does not measure end-to-end cost. The run manifest and raw responses document the reported runs, but provider-side updates and nondeterministic inference can affect replication.

%% file: sections/appendix.tex

\makeatletter
\setlength{\@fptop}{0pt}
\setlength{\@dblfptop}{0pt}
\makeatother

\section*{Guide to the Supplement}

The main paper is intentionally self-contained. This supplement provides the
additional material needed to inspect the method implementation, reproduce the
reported experiments, and trace the aggregate audit claims to concrete records.
Appendix~\ref{app:related} gives a compact lifecycle-oriented positioning table
without repeating the prose of Section~\ref{sec:related-work}.
Appendix~\ref{app:method-details} gives the notation, the complete inference
procedure with its localization and scope-selection subroutines, the serialized responsibility record, and a concrete record instance.
Appendices~\ref{app:exp-details}--\ref{app:repair-locality} document the
experimental protocol, metric definitions, and case-level repair-log schema.
Appendix~\ref{app:case-study} provides one end-to-end audit transaction.
Appendix~\ref{app:failure-analysis} reports
failure modes and applicability boundaries. Finally,
Appendix~\ref{app:boundary-prompt-map} summarizes the boundary-critical
component requirements and artifact links, and
Appendix~\ref{app:cross-domain-suites} documents the exploratory
cross-domain suites.
The main paper refers to a separate Code and Data Supplement for
machine-readable records and reproduction commands.

\section{Lifecycle-Dimension Positioning}
\label{app:related}

Section~\ref{sec:related-work} explains the conceptual distinctions in prose.
Table~\ref{tab:rw-positioning-app} provides a compact lifecycle-oriented
comparison. Its entries indicate whether a property is a required system
invariant, not whether a prior system could be extended to support it.

\begin{table*}[!t]
\centering
\scriptsize
\setlength{\tabcolsep}{3.1pt}
\renewcommand{\arraystretch}{1.12}
\begin{threeparttable}
\begin{tabularx}{\textwidth}{@{}p{0.145\textwidth}p{0.18\textwidth}YYYY@{}}
\toprule
\textbf{Method family} & \textbf{Primary addressable unit} &
\textbf{Identity allocated before code} &
\textbf{Selected code and findings retained under that identity} &
\textbf{Repair boundary as an execution constraint} &
\textbf{Persistent intervention history} \\
\midrule
Structured reasoning & Thoughts, actions, or search states &
Not generally an executable ownership requirement. &
Not generally required to bind an implementation bundle and validation record. &
Search or revision may be structured, but the boundary need not be a code-ownership invariant. &
Reasoning/search traces may be retained, but not necessarily as code-responsibility transactions. \\
\addlinespace[0.2em]
Planning and modular synthesis & Plans, function descriptions, modules, or candidate compositions &
A planning component may exist before code. &
Component identity supports synthesis, but selected candidates, runtime diagnostics, and later versions need not remain under one immutable responsibility address. &
Components may be replaced or recomposed; preservation of the outside region is not necessarily enforced as an audit invariant. &
Prior candidates or caches may be available, but append-only repair provenance is not the defining object. \\
\addlinespace[0.2em]
Execution feedback and post-hoc repair & Programs, basic blocks, files, functions, or edit spans &
Localization usually starts from the generated or existing artifact. &
Runtime diagnostics can be precise, but the diagnostic unit need not be the plan-time commitment that produced the code. &
The patch scope is selected retrospectively and may be global or artifact-derived. &
Repair iterations can be logged, but are not necessarily indexed by a pre-code responsibility identity. \\
\addlinespace[0.2em]
Traceability and formal assurance & Requirement--code links, proof obligations, specifications, or verified artifacts &
Requirements or proof objects may predate code. &
Strong property or correspondence guarantees can be retained, while the executed candidate-selection and repair history remain orthogonal. &
Verification can constrain admissible artifacts, but it does not by itself define the construction-time repair transaction. &
Depends on the workflow; construction provenance is not implied by property validity. \\
\addlinespace[0.2em]
\method & Contract-annotated responsibility node and its owned implementation bundle &
Required: the responsibility ID is allocated when the graph is created, before implementation sampling. &
Required: commitment, interface, owned code, provenance, validation findings, and versions use the same address. &
Required: repair selects an implicated node or branch, freezes the complement, records cross-boundary exceptions, or abstains. &
Required: each intervention is appended to the same responsibility record. \\
\bottomrule
\end{tabularx}
\begin{tablenotes}[flushleft]
\footnotesize
\item The distinction is lifecycle continuity: planning, generation, assembly,
validation, localization, and repair reuse one pre-code responsibility address.
The table does not claim that decomposition, feedback, traceability, or
verification are individually new.
\end{tablenotes}
\caption{Lifecycle-oriented positioning. The table complements, rather than
repeats, the related-work discussion in the main paper.}
\label{tab:rw-positioning-app}
\end{threeparttable}
\end{table*}

\section{Additional Method Details}
\label{app:method-details}

\subsection{Notation}
\label{app:notation}

Table~\ref{tab:notation} lists the symbols used in Section~\ref{sec:method} and Algorithms~\ref{alg:auditcoder}--\ref{alg:scope}. Calligraphic letters denote sets or structures; subscripts $i$ index nodes. Rule names are written in small capitals and are distinct from regions $\mathcal R$.

\begin{table*}[t]
\centering
\small
\setlength{\tabcolsep}{3pt}
\begin{tabular}{@{}p{0.35\textwidth}p{0.55\textwidth}p{0.07\textwidth}@{}}
\toprule
Symbol & Meaning & Where\\
\midrule
\multicolumn{3}{@{}l}{\textit{Inputs, outputs, and trace}}\\
$x$, $y$ & task statement; assembled program & \S\ref{sec:method}\\
$\mathcal A=(\mathcal G,\mathcal S,\mathcal E,\mathcal H)$ & audit trace: task graph, responsibility records, finding records, repair events & \S\ref{sec:method}\\
$\mathcal I=\Pi(x)=(\mathcal P,\mathcal B,\mathcal G,\mathcal Q)$ & planner output: I/O constraints, algorithmic assumptions, graph, development checks & \S\ref{subsec:contract-graph}\\
$\mathcal Q=(\chi_1,\dots,\chi_n)$; $\chi^\star$ & ordered end-to-end development checks; anchor = first failing check & \S\ref{subsec:contract-graph}, \S\ref{subsec:failure-localization}\\
\midrule
\multicolumn{3}{@{}l}{\textit{Task graph}}\\
$\mathcal G=(\mathcal V,\mathcal D_h,\mathcal D_d,\mathcal T)$ & nodes, decomposition tree, dependency DAG, order of the leaf code owners consistent with $\mathcal D_d$ & \S\ref{subsec:contract-graph}\\
$v_0$, $p(v)$, $S_h(b)$ & root; parent in $\mathcal D_h$; responsibility subtree rooted at $b$ & \S\ref{subsec:contract-graph}\\
$\mathcal U\subseteq\mathcal V$ & code-owning nodes (leaves and orchestration owners) & \S\ref{subsec:contract-graph}\\
$v_i=(\mathrm{id}_i,g_i,\tau_i,X_i,Y_i,F_i,\rho_i,C_i,q_i)$ & ID, commitment, executor category, input/output slots, contract, provenance, complexity budget, local checks & \S\ref{subsec:contract-graph}\\
$F_i$ & contract $(\mathrm{name}_i,\mathrm{args}_i,\mathrm{ret}_i,\mathrm{pre}_i,\mathrm{post}_i,\mathrm{inv}_i)$: interface fields and semantic obligations & \S\ref{subsec:contract-graph}\\
$\Theta_i$; $\mathrm{cls}(\theta)$ & obligations of $v_i$; class in $\{\acSem,\acIfc,\acRes,\acCmt\}$ & \S\ref{subsec:contract-graph}\\
$(u,v_i)$ provider pair; $\mathcal G_\rho$ & $u$ supplies an input of $v_i$ per $\rho_i$; slot-level provenance graph & \S\ref{subsec:contract-graph}\\
\midrule
\multicolumn{3}{@{}l}{\textit{Generation and findings}}\\
$\mathcal M=\{(\tau,m_\tau)\}$; $m_{\mathrm{orch}}$ & executor registry (10 algorithm generators + orchestration executor) & \S\ref{subsec:generation-assembly}\\
$a_i=\mathrm{Route}(v_i)=m_{\tau_i}$ & executor of $v_i$ (table lookup; not an owner) & \S\ref{subsec:generation-assembly}\\
$c_i$; $\mu:\mathsf{Unit}_{\mathrm{reg}}\to\mathcal U$ & owned code bundle; total ownership map over registered units & \S\ref{subsec:generation-assembly}\\
$\Omega$; $r_g=\mathrm{Execute}(y,\mathcal Q)$ & assembler; global record & \S\ref{subsec:generation-assembly}\\
$\mathrm{sym}(\chi)$ & status in $\{\textsc{Pass},\textsc{WA},\textsc{RE},\textsc{TLE},\textsc{MLE}\}$ & \S\ref{subsec:generation-assembly}\\
$e=(\mathsf{own},\mathsf{ver},\mathsf{vd},\mathsf{obl},\mathsf{wit},\mathsf{kind})$ & finding record: owners, version, verdict, obligation, witness, $\acExec$ or $\acJudg$ & \S\ref{subsec:generation-assembly}\\
$r_i$ & node-indexed findings $\{e: v_i\in e.\mathsf{own}\}$ & \S\ref{subsec:generation-assembly}\\
\midrule
\multicolumn{3}{@{}l}{\textit{Localization}}\\
$z=(W,\ell,\kappa,\widehat{\mathcal E})$ or $\bot$ & implicated set, rule $\ell\in\{\textsc{Trace},\textsc{Resource},\textsc{Local}\}$, grade $\kappa\in\{\textsc{High},\textsc{Medium},\textsc{Low}\}$, support & \S\ref{subsec:failure-localization}\\
$\Gamma(s)$ & admissible obligation classes for symptom $s$ & \S\ref{app:compat}\\
$\mathrm{Compat}(e,\chi^\star)$; $\mathcal E_c$, $\mathcal E_c^{+}$ & compatibility (C1--C6); compatible records; executed ones with observed applicability & \S\ref{subsec:failure-localization}\\
$\delta\in\{\acImpl,\acIfc,\acCmt\}$ & defect class of $z$, defined by Algorithm~\ref{alg:coherent} & \S\ref{subsec:failure-localization}\\
\midrule
\multicolumn{3}{@{}l}{\textit{Bounded repair}}\\
$\mathrm{Anc}(W)$, $\mathrm{lca}(W)$ & non-root nodes whose subtree contains $W$; its deepest element & \S\ref{subsec:bounded-repair}\\
$\mathcal R_0$; $\mathcal R^{(t)}$, $J_t$, $\mathcal J$ & nominal region; expanded region, exception set, all exceptions & App.~\ref{app:boundary}\\
$\mathcal R$; $\acRbar=\mathcal V\setminus\mathcal R$; $\partial_d\mathcal R$ & final editable scope; frozen complement; crossing dependencies & \S\ref{subsec:bounded-repair}\\
$\mathcal G',\mathcal V',\mathcal R',\mathcal U'$ & candidate graph, nodes, replacement nodes, code owners & \S\ref{subsec:bounded-repair}\\
$\mathrm{Rank}(r)$; $d$; $\sigma$; $K$ & lexicographic decision key; decision; rollback status; attempt budget & \S\ref{subsec:bounded-repair}\\
$h=(z,\mathcal R_0,\mathcal R,\acRbar,\mathcal J,\Delta\mathcal G,\Delta c,r'_g,d,\sigma)$ & repair event appended to $\mathcal H$ & \S\ref{subsec:bounded-repair}\\
\bottomrule
\end{tabular}
\caption{Notation. Symbols introduced only in Appendix~\ref{app:audit-metric-definitions} (audit metrics) are defined there.}
\label{tab:notation}
\end{table*}

\subsection{Full Inference Algorithm}
\label{app:algorithm}

Algorithm~\ref{alg:auditcoder} makes explicit the three constraints stated in
Section~\ref{sec:method}: pre-code identity allocation, ownership persistence
through assembly, and declared-scope intervention. The ownership map $\mu$ is included in
the locator call because traceback frames and registered helpers are resolved
through this map. Algorithms~\ref{alg:locate} and~\ref{alg:scope} give the
deterministic Locate and SelectScope procedures called in Stage~4, and Section~\ref{app:compat} gives the compatibility and coherence tests they use (Algorithms~\ref{alg:compat} and~\ref{alg:coherent}). We write $\mathrm{Findings}(r_g)$ for the per-check finding records emitted by an aggregate global execution result; $r_g$ remains the aggregate result supplied to localization.

\paragraph{Local findings and version eligibility.}
\label{app:local-validation-linkage}
Task-graph IR review checks planning fields and structural consistency;
it is distinct from code-level $\mathrm{VerifyLocal}$. In the reported
runs, $\mathrm{VerifyLocal}$ produces judged node-level findings; executed
evidence comes from end-to-end checks, tracebacks, and resource measurements. The word \emph{current} denotes eligibility
with respect to the active code, contract, and dependency versions, not
a particular implementation of a status flag. Findings from superseded
or rejected code versions cannot support localization of the active
program. The three-snapshot rollback audit in
Appendix~\ref{app:audit-metric-definitions} does not verify restoration
of all local-verifier state.


\begin{algorithm*}[t]
\caption{\method{} inference with attribution-guided bounded repair. Returns the final program and an audit trace with responsibility records, validation findings, and append-only repair history.}
\label{alg:auditcoder}
\small
\begin{algorithmic}[1]
\Require task $x$, executor registry $\mathcal M$, repair budget $K$
\Ensure program $y$, audit trace $\mathcal A=(\mathcal G,\mathcal S,\mathcal E,\mathcal H)$
\Statex \textit{Stage 1: allocate responsibility identities and contracts.}
\State $\mathcal I\gets\Pi(x)$; extract $(\mathcal P,\mathcal B,\mathcal G,\mathcal Q)$ \Comment{IDs and $\tau_i$ fixed here}
\State $\mathcal S,\mathcal E,\mathcal H,\mu\gets\varnothing$
\Statex \textit{Stage 2: generate node-owned code and bind local findings.}
\For{each leaf code owner $v_i$, in an order consistent with $\mathcal T$}
  \State $a_i\gets m_{\tau_i}$ \Comment{lookup in $\mathcal M$}
  \State $c_i\gets\mathrm{Generate}(a_i,g_i,F_i,X_i,Y_i,\rho_i,C_i,\mathcal B)$
  \State $\mu\gets\mathrm{Register}(\mu,v_i,c_i)$ \Comment{one owner per unit}
  \State $r_i\gets\mathrm{VerifyLocal}(c_i,F_i,q_i,C_i)$; $\mathcal E\gets\mathcal E\cup r_i$
  \State $\mathcal S\gets\mathcal S\cup\{(\mathrm{id}_i,g_i,a_i,F_i,c_i,\rho_i,r_i,\varnothing)\}$
\EndFor
\Statex \textit{Stage 3: assemble and validate.}
\State $y\gets\Omega(\mathcal G,\{(v_i,c_i)\}_{v_i\in\mathcal U})$; register orchestration units in $\mu$
  \State create/update corresponding records in $\mathcal S$ for every internal or root code owner emitted by the assembler
\State $r_g\gets\mathrm{Execute}(y,\mathcal Q)$; $\mathcal E\gets\mathcal E\cup\mathrm{Findings}(r_g)$
\Statex \textit{Stage 4: bounded repair transactions.}
\For{$k=1,\dots,K$}
  \If{$r_g$ passes all of $\mathcal Q$} \State \Return $y,\mathcal A$ \EndIf
  \State $z\gets\mathrm{Locate}(r_g,\mathcal G,\{r_i\}_{v_i\in\mathcal U},\mu)$ \Comment{Alg.~\ref{alg:locate}}
  \If{$z=\bot$} \State \Return $y,\mathcal A$ \Comment{abstain} \EndIf
  \State $\mathit{sc}\gets\mathrm{SelectScope}(z,\mathcal G,\mathcal S)$ \Comment{Alg.~\ref{alg:scope}}
  \If{$\mathit{sc}=\bot$} \State \Return $y,\mathcal A$ \Comment{no supported scope} \EndIf
  \State $(\mathcal R_0,\mathcal R,\mathcal J)\gets\mathit{sc}$; $\acRbar\gets\mathcal V\setminus\mathcal R$
  \State freeze $\{c_v,r_v\}_{v\in\acRbar\cap\mathcal U}$; $\xi_0\gets\mathrm{Snapshot}(y,\{c_v\}_{v\in\mathcal U},r_g)$
  \State $(\mathcal G'_{\mathcal R},\{c'_v\}_{v\in\mathcal R'\cap\mathcal U'})\gets\mathrm{Repair}(\mathcal G_{\mathcal R},\{c_v,r_v\}_{v\in\mathcal R\cap\mathcal U},z,\mathcal J)$
  \State $\mathcal G'\gets\mathrm{ReplaceRegion}(\mathcal G,\mathcal R,\mathcal G'_{\mathcal R})$; rebuild and check $\mu'$, $\mathcal T'$
  \State $y'\gets\Omega(\mathcal G',\{(v,c'_v)\}_{v\in\mathcal R'\cap\mathcal U'}\cup\{(v,c_v)\}_{v\in\acRbar\cap\mathcal U})$
  \If{some $c'_v\neq c_v$ with $v\in\acRbar\cap\mathcal U$, or an unrecorded edit crosses $\partial_d\mathcal R$}
    \State $d\gets\textsc{Reject}$; $r'_g\gets\varnothing$ \Comment{never enlarge $\mathcal R$ afterwards}
  \Else
    \State $r'_g\gets\mathrm{Execute}(y',\mathcal Q)$; $\mathcal E\gets\mathcal E\cup\mathrm{Findings}(r'_g)$
    \State $d\gets\textsc{Accept}$ \textbf{if} $\mathrm{Rank}(r'_g)>_{\mathrm{lex}}\mathrm{Rank}(r_g)$ \textbf{else} $\textsc{Reject}$
  \EndIf
  \If{$d=\textsc{Accept}$}
    \State $(\mathcal G,y,r_g,\mu)\gets(\mathcal G',y',r'_g,\mu')$; update records in $\mathcal R'$; $\sigma\gets\textsc{NotApplicable}$
  \Else
    \State mark candidate findings in $\mathrm{Findings}(r'_g)$ as superseded/non-current
    \State $\mathrm{Restore}(\xi_0)$; $\sigma\gets\textsc{Restored}$
  \EndIf
  \State append $h=(z,\mathcal R_0,\mathcal R,\acRbar,\mathcal J,\Delta\mathcal G,\Delta c,r'_g,d,\sigma)$ to $\mathcal H$ and to $h_v$ of the affected owners
\EndFor
\State \Return $y,\mathcal A$ \Comment{best audited state within budget}
\end{algorithmic}
\end{algorithm*}

\begin{algorithm}[t]
\caption{$\mathrm{Locate}(r_g,\mathcal G,\{r_i\},\mu)$: rule-ordered localization (\S\ref{subsec:failure-localization}). The first triggered rule decides the result; ambiguity under it yields $\bot$.}
\label{alg:locate}
\small
\begin{algorithmic}[1]
\Require failing global record $r_g$ on ordered checks $\mathcal Q$; graph $\mathcal G$; local records $\{r_i\}$; ownership map $\mu$
\Ensure $z=(W,\ell,\kappa,\widehat{\mathcal E})$ or $\bot$
\State $\chi^\star\gets$ first failing element of $\mathcal Q$; $s^\star\gets\mathrm{sym}(\chi^\star)$
\Statex \textit{Rule \textsc{Trace}}
\If{the run on $\chi^\star$ has a traceback with a generated-code frame}
  \State $f\gets$ innermost generated-code frame (harness/library frames skipped)
  \If{$\mu(u(f))$ is undefined} \State \Return $\bot$ \Comment{unresolved owner} \EndIf
  \State $\widehat{\mathcal E}\gets\{$traceback record of $\chi^\star\}$
  \State \Return $(\{\mu(u(f))\},\textsc{Trace},\textsc{High},\widehat{\mathcal E})$
\EndIf
\Statex \textit{Rule \textsc{Resource}}
\State $O\gets\{v\in\mathcal U:$ an executed measurement attributed to $v$ via $\mu$ violates $C_v\}$
\If{$s^\star\in\{\textsc{TLE},\textsc{MLE}\}$ \textbf{and} $O\neq\varnothing$}
  \If{$|O|>1$} \State \Return $\bot$ \EndIf
  \State $\widehat{\mathcal E}\gets$ the measurements of the owner in $O$
  \State \Return $(O,\textsc{Resource},\textsc{Medium},\widehat{\mathcal E})$
\EndIf
\Statex \textit{Rule \textsc{Local}}
\State $\mathcal E_c\gets\{\mathrm{Compat}(e,\chi^\star):e\in\bigcup_i r_i\}\setminus\{\bot\}$ \Comment{Alg.~\ref{alg:compat}; incl.\ (C5) re-addressing}
\If{$\mathcal E_c\neq\varnothing$}
  \State $\mathcal E_c^{+}\gets\{e\in\mathcal E_c: e.\mathsf{kind}=\acExec$, applicability observed$\}$
  \If{$\mathcal E_c^{+}\neq\varnothing$} \State $(\widehat{\mathcal E},\kappa)\gets(\mathcal E_c^{+},\textsc{Medium})$
  \Else \State $(\widehat{\mathcal E},\kappa)\gets(\mathcal E_c,\textsc{Low})$
  \EndIf
  \State $W\gets\bigcup_{e\in\widehat{\mathcal E}}e.\mathsf{own}$
  \If{$\neg\,\mathrm{Coherent}(\widehat{\mathcal E},W)$} \State \Return $\bot$ \Comment{Alg.~\ref{alg:coherent}; competing explanations} \EndIf
  \State \Return $(W,\textsc{Local},\kappa,\widehat{\mathcal E})$
\EndIf
\State \Return $\bot$ \Comment{no rule triggered}
\end{algorithmic}
\end{algorithm}

\begin{algorithm}[t]
\caption{$\mathrm{SelectScope}(z,\mathcal G,\mathcal S)$: nominal region and final editable scope (\S\ref{subsec:bounded-repair}).}
\label{alg:scope}
\small
\begin{algorithmic}[1]
\Require $z=(W,\ell,\kappa,\widehat{\mathcal E})\neq\bot$; graph $\mathcal G$; records $\mathcal S$
\Ensure $(\mathcal R_0,\mathcal R,\mathcal J)$ or $\bot$
\State $\delta\gets\mathrm{DefectClass}(\widehat{\mathcal E})$ \Comment{Alg.~\ref{alg:coherent}}
\If{$W=\{v\}$ \textbf{and} $v\in\mathcal U$ \textbf{and} $\delta=\acImpl$}
  \State $\mathcal R_0\gets\{v\}$ \Comment{change $c_v$ under fixed $F_v$}
\Else
  \State $\mathcal R_0\gets\bot$
  \For{$b\in\mathrm{Anc}(W)$, deepest first}
    \If{($\delta=\acCmt$ and $\widehat{\mathcal E}$ names $g_b$) \textbf{or} ($\delta=\acIfc$ and all named provider pairs lie in $S_h(b)$)}
      \State $\mathcal R_0\gets S_h(b)$; \textbf{break}
    \EndIf
  \EndFor
  \If{$\mathcal R_0=\bot$} \State \Return $\bot$ \Comment{no supported local boundary} \EndIf
\EndIf
\State $\mathcal R\gets\mathcal R_0$; $\mathcal J\gets\langle\rangle$
\While{an intended edit lies outside $\mathcal R$ or changes a pair in $\partial_d\mathcal R$}
  \State $J_t\gets$ owners declared by the next integration exception
  \If{the exception lacks recorded supporting findings} \State \Return $\bot$ \EndIf
  \If{it requests structural replanning} \State $J_t\gets\bigcup_{u\in J_t}S_h(u)$ \EndIf
  \State $\mathcal R\gets\mathcal R\cup J_t$; append the exception to $\mathcal J$
\EndWhile
\If{$\mathcal R=\mathcal V$ \textbf{or} $(\mathcal V\setminus\mathcal R)\cap\mathcal U=\varnothing$} \State \Return $\bot$ \EndIf
\State \Return $(\mathcal R_0,\mathcal R,\mathcal J)$
\end{algorithmic}
\end{algorithm}

\subsection{Compatibility, Coherence, and Boundary Expansion}
\label{app:compat}

This section specifies the compatibility, coherence, and evidence-grading
rules summarized in Section~\ref{subsec:failure-localization}.
Algorithms~\ref{alg:compat} and~\ref{alg:coherent} give the tests used by
localization and scope selection.

\paragraph{Full compatibility predicate.}
Obligations are classified as semantic ($\acSem$), interface/provenance
($\acIfc$), resource ($\acRes$), or planning commitment ($\acCmt$).
The admissible classes for each failure symptom are
\[
\begin{aligned}
\Gamma(\textsc{WA})&=\Gamma(\textsc{RE})
 =\{\acSem,\acIfc,\acCmt\},\\
\Gamma(\textsc{TLE})&=\Gamma(\textsc{MLE})
 =\{\acRes,\acCmt\}.
\end{aligned}
\]
For a finding $e$, let $\theta=e.\mathsf{obl}$ and
$s^\star=\mathrm{sym}(\chi^\star)$. Compatibility is determined by
conditions (C1)--(C6):
\emph{(C1) currency and specificity:} $e$ is current,
$e.\mathsf{vd}=\textsc{Reject}$, and $\theta\ne\varnothing$;
\emph{(C2) symptom consistency:} $\mathrm{cls}(\theta)\in\Gamma(s^\star)$;
\emph{(C3) reachability:} $\mathcal G_\rho$ contains a path from a
slot constrained by $\theta$ to the observable failing on $\chi^\star$;
\emph{(C4) applicability:} the anchor input satisfies the input
condition recorded in $e.\mathsf{wit}$;
\emph{(C5) precondition scope:} when the witness input of a single-owner
record violates $\mathrm{pre}_v$, the record cannot support an
implementation defect of $v$ and is re-addressed as an $\acIfc$ record
on the provider pair $(u,v)$ named by $\rho_v$, with
$e.\mathsf{own}=\{u,v\}$;
\emph{(C6) non-contradiction:} no current executed record of the same
owner passes $\theta$ under the same input condition.
Algorithm~\ref{alg:compat} applies (C5) first, followed by
(C1)--(C4) and (C6).

Applicability is \emph{observed} when the anchor execution records a
call to an owned unit whose arguments satisfy the witness condition;
it is \emph{asserted} when supported only by a verifier statement.
An asserted condition does not constitute an execution witness.

\begin{algorithm}[t]
\caption{$\mathrm{Compat}(e,\chi^\star)$: compatibility conditions (C1)--(C6). Returns the possibly re-addressed record or $\bot$.}
\label{alg:compat}
\small
\begin{algorithmic}[1]
\Require local record $e$; anchor $\chi^\star$ with symptom $s^\star$
\Ensure $e$ (or its re-addressed copy) or $\bot$
\If{$e$ has one owner $v$ \textbf{and} its witness input violates $\mathrm{pre}_v$} \Comment{(C5)}
  \State $u\gets$ provider of that input per $\rho_v$
  \State $e\gets e$ with $e.\mathsf{own}=\{u,v\}$ and $e.\mathsf{obl}=$ the $\acIfc$ obligation of $(u,v)$
\EndIf
\State $\theta\gets e.\mathsf{obl}$
\If{$e$ not current \textbf{or} $e.\mathsf{vd}\neq\textsc{Reject}$ \textbf{or} $\theta=\varnothing$} \State \Return $\bot$ \Comment{(C1)} \EndIf
\If{$\mathrm{cls}(\theta)\notin\Gamma(s^\star)$} \State \Return $\bot$ \Comment{(C2)} \EndIf
\If{$\mathcal G_\rho$ has no path from a slot constrained by $\theta$ to the failing observable} \State \Return $\bot$ \Comment{(C3)} \EndIf
\If{$\chi^\star$ does not satisfy the input condition of $e.\mathsf{wit}$} \State \Return $\bot$ \Comment{(C4)} \EndIf
\If{a current executed record of the same owner passes $\theta$ under the same input condition} \State \Return $\bot$ \Comment{(C6)} \EndIf
\State \Return $e$
\end{algorithmic}
\end{algorithm}

\paragraph{Coherence and defect class.}
Algorithm~\ref{alg:coherent} defines coherence and the defect class.
When commitment obligations are present, they must all name the same
$g_b$ and satisfy $W\subseteq S_h(b)$. Otherwise, coherence requires
$|W|=1$ or a connected graph on $W$ formed by the named interface
provider pairs. Non-coherent records constitute competing explanations,
and localization abstains rather than merging them into a wider region.
The defect class prioritizes $\acCmt$, then $\acIfc$, and otherwise
returns $\acImpl$. The last case includes \textsc{Trace} and
\textsc{Resource} results whose records name no obligation.

\begin{algorithm}[t]
\caption{$\mathrm{Coherent}(\widehat{\mathcal E},W)$ and $\mathrm{DefectClass}(\widehat{\mathcal E})$: coherence and defect classification.}
\label{alg:coherent}
\small
\begin{algorithmic}[1]
\Function{Coherent}{$\widehat{\mathcal E},W$}
  \State $\Theta_{\mathrm{c}}\gets$ the $\acCmt$ obligations named by $\widehat{\mathcal E}$
  \If{$\Theta_{\mathrm{c}}\neq\varnothing$}
    \State \Return [all of $\Theta_{\mathrm{c}}$ name one commitment $g_b$ \textbf{and} $W\subseteq S_h(b)$] \Comment{shared commitment}
  \Else
    \State \Return [$|W|=1$ \textbf{or} the graph on $W$ with one edge per provider pair named by $\acIfc$ records of $\widehat{\mathcal E}$ is connected] \Comment{interaction}
  \EndIf
\EndFunction
\Function{DefectClass}{$\widehat{\mathcal E}$}
  \If{$\widehat{\mathcal E}$ names a $\acCmt$ obligation} \State \Return $\acCmt$
  \ElsIf{$\widehat{\mathcal E}$ names an $\acIfc$ obligation} \State \Return $\acIfc$
  \Else \State \Return $\acImpl$ \EndIf
\EndFunction
\end{algorithmic}
\end{algorithm}

\paragraph{Attribution grades.}
The grade $\kappa$ is recomputed from stored evidence. \textsc{High} requires a generated-code traceback frame resolved through $\mu$ in the failing execution. \textsc{Medium} requires an executed finding or owner-scoped resource measurement with observed applicability. \textsc{Low} is used for a judged rejection or an executed finding whose applicability is only asserted. Grades are ordinal evidence labels rather than calibrated probabilities and do not alter scope selection.

\begin{table}[t]
\centering
\small
\setlength{\tabcolsep}{4pt}
\begin{tabular}{@{}lp{0.73\columnwidth}@{}}
\toprule
$\kappa$ & Evidence condition \\
\midrule
\textsc{High} & Generated-code frame of the failing run, resolved through $\mu$; same execution; \textsc{Trace}.\\
\textsc{Medium} & Executed finding or measured cost attributed through $\mu$; applicability observed; \textsc{Resource} or \textsc{Local}.\\
\textsc{Low} & Judged rejection, or executed finding whose applicability is only asserted; \textsc{Local}.\\
\bottomrule
\end{tabular}
\caption{Attribution grades used by localization. A grade records the strength of stored evidence and is not model confidence.}
\label{tab:grades}
\end{table}

\paragraph{Boundary expansion.}
\label{app:boundary}
Starting from $\mathcal R^{(0)}=\mathcal R_0$, each approved integration exception adds its explicitly declared affected set $J_t$:
\[
\begin{aligned}
 \mathcal R^{(t+1)}&=\mathcal R^{(t)}\cup J_t,\\
 \mathcal R&=\mathcal R^{(t_*)}.
\end{aligned}
\]
An exception records the affected owner IDs, the intended code, contract, or structure edits, and the supporting findings, all before regeneration. Scope membership, hierarchy descendants, and exception owner IDs are resolved in the active graph before this repair transaction replaces any nodes; this is also the pre-repair snapshot used for RRS in Appendix~\ref{app:audit-metric-definitions}. A structural-replanning request adds the entire affected hierarchy subtree, $J_t\leftarrow\bigcup_{u\in J_t}S_h(u)$. The loop in Algorithm~\ref{alg:scope} runs until every intended edit lies inside $\mathcal R$ and no pair in $\partial_d\mathcal R$ changes. An exception without recorded findings, $\mathcal R=\mathcal V$, or a scope with no frozen code owner returns $\bot$ and ends bounded repair. A shared helper outside $\mathcal R$ stays frozen; a helper inside $\mathcal R$ that is consumed outside must retain every exposed boundary contract, not only the interface of the branch root. Boundary checks test these requirements but do not prove semantic equivalence.

\subsection{Lifecycle Invariants and Serialized Audit Schema}
\label{app:schema}

For code-owning node $v_i$, the conceptual responsibility record is
\[
  s_i=(\mathrm{id}_i,g_i,a_i,F_i,c_i,\rho_i,r_i,h_i).
\]
The serialized form must therefore contain the owned implementation bundle
explicitly; omitting code from the schema would break the decision--code link
that the method claims to preserve. We use the following record:
\[
\begin{gathered}
\mathrm{Node}_i=\{\\
\text{\normalfont\ttfamily id},
\text{\normalfont\ttfamily goal},
\text{\normalfont\ttfamily parent},\\
\text{\normalfont\ttfamily dependencies},
\text{\normalfont\ttfamily agent\_type},\\
\text{\normalfont\ttfamily interface},
\text{\normalfont\ttfamily complexity\_budget},\\
\text{\normalfont\ttfamily owned\_code},
\text{\normalfont\ttfamily provenance},\\
\text{\normalfont\ttfamily validation},\\
\text{\normalfont\ttfamily integration\_exceptions},\\
\text{\normalfont\ttfamily repair\_history}\}.
\end{gathered}
\]

\begin{table*}[!t]
\centering
\small
\setlength{\tabcolsep}{3.8pt}
\renewcommand{\arraystretch}{1.1}
\begin{tabularx}{\textwidth}{@{}p{0.20\textwidth}p{0.36\textwidth}Y@{}}
\toprule
\textbf{Field group} & \textbf{Stored content} & \textbf{Lifecycle rule} \\
\midrule
Identity and structure & \texttt{id}, \texttt{goal}, \texttt{parent}, and
\texttt{dependencies}. & \texttt{id} is allocated when the graph is created,
before any implementation is sampled, and is not reassigned after repair. \\
Agent routing & \texttt{agent\_type} and the concrete routed agent identifier. &
The agent answers \emph{who executes}; it does not replace the responsibility ID
that answers \emph{which code and findings are owned}. \\
Interface contract & Function name, typed parameters, return schema,
preconditions, postconditions, invariants, and complexity budget. & Semantic clauses
are checked by the LLM verifier unless instantiated as tests or assertions;
judged findings carry no execution witness. \\
Owned implementation & \texttt{owned\_code}: the selected function/helper bundle
registered under the node. & The ownership map $\mu$ is total over registered code
units and maps each unit to exactly one responsibility owner. \\
Provenance and findings & Variable-source map, local tests, contract checks,
resource probes, and global findings linked back to the node. & Code and findings
reuse the same address; a passing record means accepted by a test or by the
verifier, not formally proved. \\
Exceptions and history & Cross-node integration exceptions, selected region,
frozen complement, graph/code deltas, validation outcome, and prior versions. &
Cross-boundary edits must be explicit and expand the transaction boundary;
repair history is append-only. \\
\bottomrule
\end{tabularx}
\caption{Serialized responsibility-record fields and the invariants they support.}
\label{tab:serialized-schema}
\end{table*}

\subsection{Concrete Responsibility-Record Instance}
\label{app:responsibility-graph-instance}

\begin{figure*}[!t]
    \centering
    \includegraphics[width=0.93\textwidth]{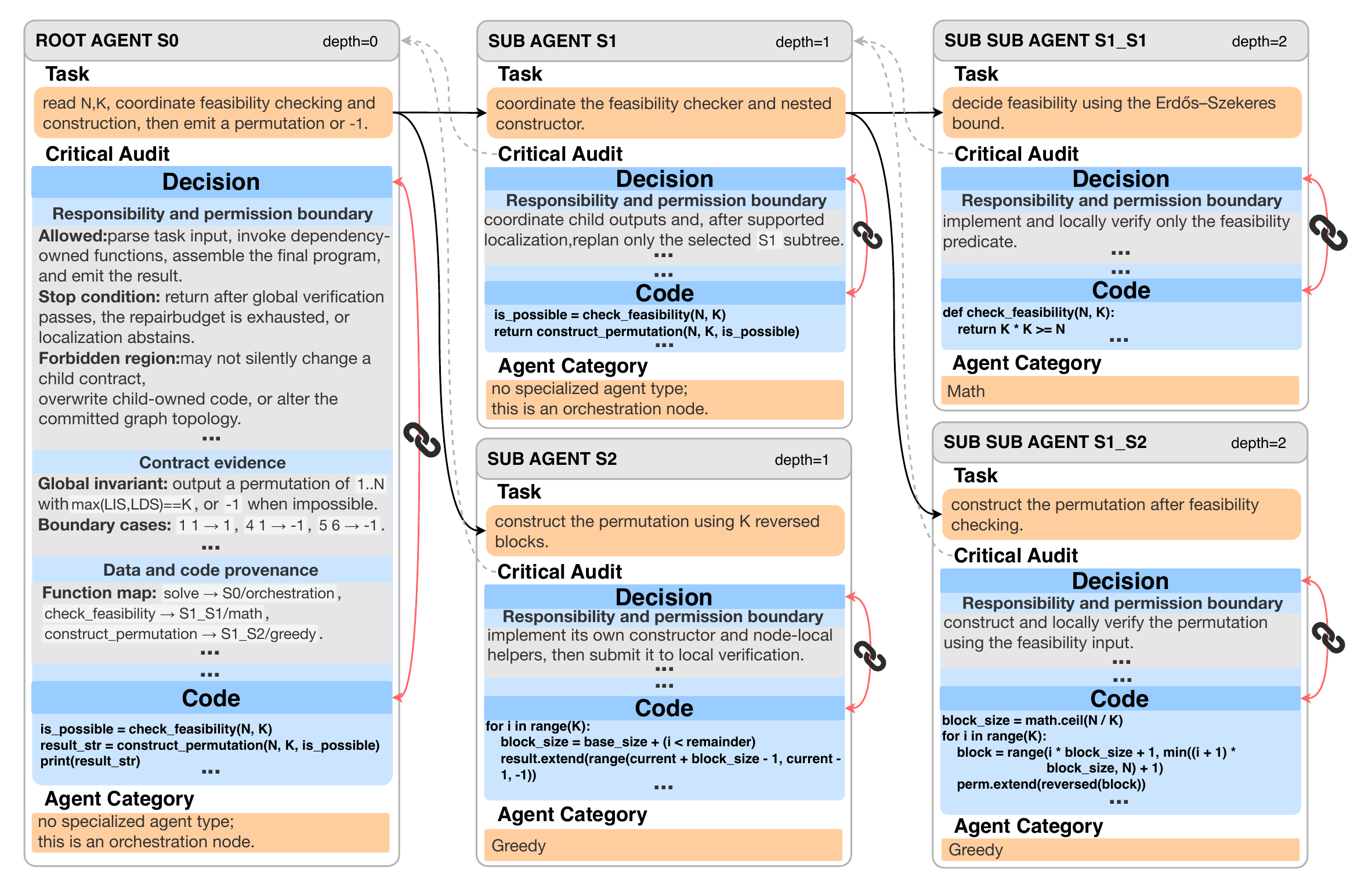}
    \caption{Task breakdown and critical-audit records for the
permutation-construction example. Solid black arrows show task dependencies
fixed by the global planner before code generation; gray dashed arrows show
child-to-parent code and audit-record propagation for assembly; red
bidirectional arrows mark the recorded decision--code link within each node.
Agent headings denote routed execution roles, while responsibility records
remain the addressable audit units. The partial code illustrates record
structure and is not evidence that the completed program satisfies the
global contract.}
    \label{fig:full-responsibility-graph}
\end{figure*}

Figure~\ref{fig:full-responsibility-graph} instantiates the serialized schema
with the permutation-construction example. It presents the task breakdown across root, child, and grandchild
responsibility records together with the critical-audit fields retained
for each unit. The agent-oriented headings identify
routed execution roles; the node identities and the responsibility records inside
the cards remain the addressable units used by \method. The figure is a
static artifact snapshot; Figure~\ref{fig:auditcoder-lifecycle} in the main paper
shows how the records are used over time for validation, localization, and bounded
repair.

The distinction between an agent role and a responsibility address is essential.
For example, ``Greedy'' identifies a routed executor, whereas the node identity is
the stable lookup key for its commitment, selected code, validation findings, and
intervention history. The function map displayed in \texttt{S0} assigns
\texttt{check\_feasibility} to \texttt{S1\_S1} and
\texttt{construct\_permutation} to \texttt{S1\_S2}. The parallel constructor
shown under \texttt{S2} therefore exposes a duplicated responsibility that must be
marked unused or resolved by the ownership checker before assembly; it is not a
second valid owner. The parent consumes the mapped child records during assembly;
it does not acquire ownership of the child code.

\paragraph{Scope of the displayed code.}
The visible fragments are insufficient to establish the stated
$\max(\mathrm{LIS},\mathrm{LDS})=K$ contract. By itself,
\texttt{K * K >= N} accepts $(N,K)=(5,6)$, contrary to the displayed
infeasible boundary case. For $(N,K)=(10,6)$, the fixed
$\lceil N/K\rceil$-sized blocks shown in \texttt{S1\_S2} produce
\[
[2,1,4,3,6,5,8,7,10,9],
\]
whose LIS and LDS lengths are 5 and 2, respectively. These observations
concern only the displayed fragments; omitted code and the validation
status of the full program are not established by this figure.


\section{Experimental Protocol and Reproducibility}
\label{app:exp-details}

This section records the operational details needed to reproduce the functional
and audit analyses. Aggregate outcomes are reported in
Tables~\ref{tab:functional-results}, \ref{tab:audit-metrics},
and~\ref{tab:injection-repair} of the main paper.

\begin{table*}[!t]
\centering
\small
\setlength{\tabcolsep}{3.8pt}
\renewcommand{\arraystretch}{1.09}
\begin{threeparttable}
\begin{tabularx}{\textwidth}{@{}p{0.18\textwidth}p{0.29\textwidth}Y@{}}
\toprule
\textbf{Item} & \textbf{Evaluated records} & \textbf{Protocol and disclosure} \\
\midrule
Primary APPS functional runs & Fixed 200-task sample: 50 introductory, 100 interview,
and 50 competition tasks. & The same problem statements, execution interface,
timeout, memory policy, and decoding configuration are used across methods.
Complete methods receive at most one feedback-conditioned debugging, retry,
or repair round ($K{=}1$); no-repair variants receive none. \\
ClassEval functional runs & 100 class-level records per model--method run. &
The class-level harness checks interdependent methods and shared state; the
denominator is fixed at 100 for each reported row, including the
component configurations in Table~\ref{tab:functional-results}. \\
Audit records & The 200 APPS records (50 introductory, 100 interview, and 50
competition tasks) of the \texttt{deepseek-v4-flash} \method{} functional run. &
Table~\ref{tab:audit-metrics} is computed read-only from the construction
traces of that run; its 52 initially failing tasks and 30 passing repairs
reconcile with the 89.0\% \texttt{pass@1} in Table~\ref{tab:functional-results}. \\
Repair subset & 52 initially failing tasks: 44 contain a recorded node/branch
localization; for the remaining 8, the locator abstains, no repair call is
made, and the task counts as unrepaired. & A transaction is a bounded repair
attempt with a recorded accept or rollback decision. Localized repair success,
RRS, CCR, and decision compliance use the 44 localized transactions; rollback
integrity uses their 14 rollbacks. \\
Controlled fault injection & The 100 interview tasks of the APPS sample;
four fault classes of 25 tasks each (runtime exception, wrong answer, interface
contract, resource timeout). & For each task and method, a passing program is
selected from several samples of the method's own pipeline and checked; the
same fault is then injected into the code unit that method holds responsible.
The fault marker is never visible to any method. Each method is scored after its
first feedback-conditioned round. Post-injection local checks are rerun
to obtain current findings but excluded from the reported repair-round
cost for all methods. AgentCoder counts only its programmer call.
Abstentions make no repair call; per-task averages still use all 100 tasks. \\
Model and call metadata & \texttt{deepseek-v4-flash} and Qwen3.6-plus for the
functional tables; the audit uses the traces of the \texttt{deepseek-v4-flash}
APPS run. & The run manifest records the provider-visible model, input files,
and file hashes. \\
Sandbox records & Every executed program. & The harness records exit status,
stdout, stderr, traceback frames, observed and expected outputs, runtime,
timeout status, and memory status when available. \\
\bottomrule
\end{tabularx}
\caption{Evaluation denominators, exclusions, and logged metadata. Exact numeric
model settings and sandbox limits should be read from the run manifest included
in the Code and Data Supplement, not inferred from aggregate tables.}
\label{tab:eval-config}
\end{threeparttable}
\end{table*}

\paragraph{Baseline operations.}
AgentCoder~\citep{huang2023agentcoder} uses programmer, test-designer, and
test-executor roles with iterative execution feedback.
MapCoder~\citep{islam2024mapcoder} separates example recall, planning, coding,
and debugging into four agents and ranks candidate plans by model-generated
confidence. CodeSim~\citep{islam-etal-2025-codesim} adds plan verification and
internal debugging through LLM-simulated input/output traces. Each baseline uses
the same base model, task inputs, and execution harness as \method.
In the primary setting, its debugging or retry loop uses at most one
feedback-conditioned round, matching the single-intervention budget
($K{=}1$) of \method. In the controlled
fault-injection study, each baseline runs its native retry procedure, and its
result is taken after the first round.
\paragraph{Component configurations.}
On APPS and ClassEval, Table~\ref{tab:functional-results} reports the
same three component labels: \emph{Task plan}, \emph{+ algorithm agents},
and \emph{\method{} (+ repair)}. They denote task-graph planning,
the addition of algorithm-specialized generators, and the further
addition of attribution-guided bounded repair, respectively.
The first two configurations receive no repair round; the complete
pipeline receives at most one in the primary setting.


\paragraph{Run selection and denominators.}
Fixed-run \texttt{pass@1} scores one retained output per evaluated record.
In the primary functional setting, hidden-test outcomes are used only
for evaluation, not candidate selection. Rows with debugging, retry,
or repair consume at most one such round per task; no-repair variants
receive none, and abstention makes no repair call. This protocol is
distinct from selecting initially passing programs to construct the
controlled fault-injection cohort.

\paragraph{Adaptive multi-round APPS runs.}
\label{app:multiround-stopping}
The final block of Table~\ref{tab:functional-results} evaluates the four
complete methods without a fixed round cap. Improvement means a strict
increase in the number of passed tests. An improving round resets the
non-improvement counter; an unchanged or lower pass count does not.
The stopping rule is four consecutive non-improving rounds. This is a
stopping criterion, not a shared candidate-acceptance policy: it does
not replace a baseline's own update rule with \method{}'s lexicographic
$\mathrm{Rank}$. These runs are excluded from the process audit in
Table~\ref{tab:audit-metrics} and reported separately from the primary
single-intervention setting.


\paragraph{Fault injection and manual scope inspection.}
In the controlled cohort, one initially hidden-test-passing program is
selected per task and method before a semantically matched fault is
injected. Injection markers are bookkeeping only and are hidden from
all methods. For non-abstaining \method{} tasks, manual inspection
checks whether the selected scope contains the responsibility owner
of the injected fault. The exact inspected scope is denoted by
$\mathcal R_t^{\mathrm{checked}}$ in
Appendix~\ref{app:audit-metric-definitions}; containment is evaluated
separately from post-repair hidden-test success. The logical
case-to-scope linkage is described in
Appendix~\ref{app:repair-locality}.

\paragraph{Repair-round cost accounting.}
Token counts in Table~\ref{tab:injection-repair} cover the first
feedback-conditioned repair round, excluding post-injection local
checks. Per-task averages use all 100 injected tasks, including
abstentions. Per-fix averages divide all counted repair tokens,
including unsuccessful attempts, by the number of successful fixes.
For AgentCoder, only the programmer call is counted; its test-designer
call is excluded. These figures are not end-to-end pipeline costs.

\paragraph{Abstention.}
A locator or scope output of $\bot$ terminates the repair procedure in
Algorithm~\ref{alg:auditcoder} without a repair call; the assembled program is
returned unchanged and scored as is. In the audited run, this applies to 8 of
the 52 initially failing tasks. They count as failures in \texttt{pass@1} and
are excluded from localized-repair success, RRS, CCR, and the governance
metrics.

\paragraph{Reproducibility materials.}
This appendix documents the method specification, experimental cohorts,
metric definitions, and case records; verbatim prompt templates are not
reproduced. The main paper
refers separately to implementation artifacts and reproduction commands
in the Code and Data Supplement. Exact model settings, sandbox limits,
and artifact versions cannot be inferred from aggregate tables.


\section{Audit Metric Definitions}
\label{app:audit-metric-definitions}

Let $\mathcal{N}$ be the set of audited task records. For the structural
metrics and $t\in\mathcal{N}$, let
$\mathcal{G}_t=(\mathcal{V}_t,\mathcal{D}_{h,t},\mathcal{D}_{d,t},\mathcal{T}_t)$ be its
final task graph in the notation of Section~\ref{subsec:contract-graph},
$\mathcal{U}_t\subseteq\mathcal{V}_t$ its code-owning nodes, and
$\mathcal{L}_t\subseteq\mathcal{U}_t$ its leaf code owners; let
$\mathcal{U}=\bigcup_{t\in\mathcal{N}}\mathcal{U}_t$ denote their pooled set.
Repair transactions are the events $h\in\mathcal H$ of these records.

\paragraph{Decision--code trace coverage.}
For leaf node $v_i$, let $c_i$, $F_i$, $\rho_i$, and $r_i$ denote its owned code,
basic interface contract, input-source fields, and validation record. Then
\[
\begin{aligned}
\mathrm{DCTC}_{t}&=
\frac{1}{|\mathcal{L}_t|}\sum_{v_i\in\mathcal{L}_t}\\
&\quad\mathbb{I}[c_i\neq\emptyset\wedge F_i\neq\emptyset\\
&\qquad\wedge\rho_i\neq\emptyset\wedge r_i\neq\emptyset].
\end{aligned}
\]
Table~\ref{tab:audit-metrics} reports the task-macro average
$|\mathcal{N}|^{-1}\sum_{t\in\mathcal{N}}\mathrm{DCTC}_t$. Initial DCTC is computed after
initial assembly; final DCTC is computed after repair and reassembly. A
nonempty verdict counts even when it is \texttt{reject}; DCTC therefore measures
record presence, not code correctness, provenance truth, or contract execution.

\paragraph{Graph integrity.}
$\mathrm{GraphValid}(t)$ requires nonempty unique node IDs, a resolvable root,
valid parent and dependency references, an acyclic parent relation, no duplicate
execution-order entries, and equality between the execution-order set and the
recorded leaf set. The graph-integrity rate is
\[
\mathrm{GIR}=
\frac{1}{|\mathcal{N}|}\sum_{t\in\mathcal{N}}
\mathbb{I}[\mathrm{GraphValid}(t)].
\]
This is a post-hoc structural check. It does not establish that a decomposition
or dependency edge is semantically correct.

\paragraph{Contract-signature consistency.}
For code-owning node $v_i$, including orchestration owners,
$\mathrm{SigMatch}(v_i)$ requires parseable Python code whose owned function
has positional parameter names and order that match the declared interface
after an optional leading \texttt{self} is removed; when the declared
interface names a function, the code must also contain exactly one function
with that name. The declared
interface of a leaf code owner is the one produced by the interface designer;
that of an orchestration owner is the input/output slot list declared for it by
the task decomposer. Then
\[
\mathrm{CSCR}=
\frac{\sum_{v_i\in\mathcal{U}}\mathbb{I}[\mathrm{SigMatch}(v_i)]}
{|\mathcal{U}|}.
\]
The check does not cover type annotations, default or keyword-only arguments,
preconditions, postconditions, invariants, complexity, or functional behavior.

\paragraph{Recorded localization coverage and localized repair success.}
Let $\mathcal{F}_{\mathrm{repair}}$ be initially failed tasks admitted to repair,
and let $\mathrm{RecordedLocate}(t)$ indicate that the persisted record names a
node or branch and records a localization method. Define
$\mathcal{F}_{\mathrm{loc}}=\{t\in\mathcal{F}_{\mathrm{repair}}:
\mathrm{RecordedLocate}(t)\}$. Then
\[
\begin{aligned}
\mathrm{LocCov}
&=\frac{|\mathcal{F}_{\mathrm{loc}}|}
{|\mathcal{F}_{\mathrm{repair}}|},\\
\mathrm{LRS}_{\mathrm{APPS}}
&=\frac{|\{t\in\mathcal{F}_{\mathrm{loc}}:
\mathrm{Pass}_{\mathrm{APPS}}(\mathrm{Repair}(t))\}|}
{|\mathcal{F}_{\mathrm{loc}}|}.
\end{aligned}
\]
In the audited run, 44 of the 52 initially failing tasks have a recorded
localization; for the remaining eight, the locator abstains and no repair is
attempted.

\paragraph{Fault-owner containment under injection.}
\label{app:fault-owner-containment}
Let $\mathcal N_{\mathrm{inj}}$ be the injected tasks and
$\mathcal N_{\mathrm{scope}}\subseteq\mathcal N_{\mathrm{inj}}$ those
receiving a repair scope. For task $t$, let $v_t^{\mathrm{fault}}$ be
the owner of the injected fault and $\mathcal R_t^{\mathrm{checked}}$
the scope used in manual inspection. Define
\[
\begin{aligned}
I_t^{\mathrm{FOC}}
 &=\mathbb I[v_t^{\mathrm{fault}}\in\mathcal R_t^{\mathrm{checked}}],\\
\mathrm{FOC}
 &=\frac{\sum_{t\in\mathcal N_{\mathrm{scope}}}I_t^{\mathrm{FOC}}}
 {|\mathcal N_{\mathrm{scope}}|}.
\end{aligned}
\]
The reported rate is $88/93$ (94.6\%). Five selected scopes do not
contain the injected fault owner. Including the seven abstentions
as non-contained tasks gives $88/100$ (88.0\%). Containment and
repair success are evaluated separately: membership in the selected
scope does not establish that the replacement passes the tests.
The fault owner and inspected scope must use the same graph and
ownership version; $\mathcal R_t^{\mathrm{checked}}$ denotes the
inspected object without assuming that it is $\mathcal R_0$ or the
expanded $\mathcal R$.


\paragraph{Repair region and changed code.}
\label{app:rrs-snapshot}
For a localized transaction $h$, let $\mathcal G_h^{\mathrm{pre}}$
be the active task graph immediately before repair, with node set
$\mathcal V_h^{\mathrm{pre}}$. Let
$\mathcal R_h\subseteq\mathcal V_h^{\mathrm{pre}}$ be the final declared
editable scope, including all recorded integration expansions before
regeneration. Both sets refer to the same pre-repair node-ID space.
For programs $y_h$ and $y'_h$ before and after the persisted repair
decision, define
\[
\begin{aligned}
\mathrm{RRS}_{\mathrm{nodes}}(h)
 &=\frac{|\mathcal R_h|}{|\mathcal V_h^{\mathrm{pre}}|},\\
\mathrm{CCR}_{\mathrm{LOC}}(h)
 &=\frac{\mathrm{ChangedLOC}(y_h,y'_h)}{\mathrm{LOC}(y'_h)}.
\end{aligned}
\]
RRS measures the fraction of pre-existing responsibility nodes that the
transaction authorizes for revision. Its denominator includes all graph
nodes, not only leaf or code-owning nodes. Branch membership and any
expansion are resolved in $\mathcal G_h^{\mathrm{pre}}$; fresh replacement
IDs belong to the candidate graph and do not alter this transaction's
RRS. This prevents splitting or merging nodes during replanning from
changing the measurement of the original permission boundary.

RRS is recorded for both accepted and rolled-back transactions: rollback
does not erase the declared scope, although the persisted code delta is
zero. CCR instead measures the retained textual change. Its reported
values mix accepted repairs and rollback-induced zero deltas and are
not the delta distribution of accepted candidates alone. The scope
statistics in Table~\ref{tab:audit-metrics} summarize the same 44
localized transactions. In a multi-round run, each transaction would
use its own immediately preceding active graph, not the run's final
graph or a union of scopes; such runs are excluded from that table.


\paragraph{Test regression.}
For $t\in\mathcal F_{\mathrm{loc}}$, let $\mathrm{HidPass}(y_t)$ be the set of
hidden APPS tests passed by the program $y_t$ before localized repair, and let
$y'_t$ be the program after the persisted repair decision. Define
$\mathcal F_t=\{\eta\in\mathrm{HidPass}(y_t):
\eta\text{ fails on }y'_t\}$. Pooling over the 44 localized tasks,
\[
\mathrm{Regress}
 =\frac{\displaystyle\sum_t|\mathcal F_t|}
 {\displaystyle\sum_t|\mathrm{HidPass}(y_t)|}.
\]
The main-table denominator is 131 eligible test observations, none of which
fails after repair; it is not the number of repaired tasks.

\paragraph{Strict-improvement decision compliance.}
The recorded candidate ordering is the lexicographic key $\mathrm{Rank}(r)$ of
Section~\ref{subsec:bounded-repair}, whose components indicate full pass,
external passes, internal passes, external clean exits, and internal clean
exits; ``external'' and ``internal'' refer to the task's own examples and the
executed planner-derived boundary cases in $\mathcal Q$, respectively. A
candidate is expected to be accepted exactly when its rank exceeds the
pre-repair rank. For transaction $h$ with pre-repair record $r_g$ and candidate
record $r'_g$, let $d_h$ be the recorded binary decision and
$\hat d_h=\mathbb{I}[\mathrm{Rank}(r'_g)>_{\mathrm{lex}}\mathrm{Rank}(r_g)]$ the
expected decision. For the transactions $\mathcal{H}_{\mathrm{rec}}$ with
acceptance metadata,
\[
\mathrm{SIDC}=
\frac{1}{|\mathcal{H}_{\mathrm{rec}}|}
\sum_{h\in\mathcal{H}_{\mathrm{rec}}}
\mathbb{I}[d_h=\hat d_h].
\]
A transaction whose candidate is rejected by the boundary check before
execution has no candidate rank; its expected decision is rejection. This
checks policy consistency against recorded ranks, not whether the tests or
ranks are themselves complete. All 44 localized transactions carry acceptance
records.

\paragraph{Rollback integrity.}
For rollback transaction $h$, $\mathrm{Restored}(h)$ requires equality between
the final and initial values of (i) assembled code, (ii) node code snippets, and
(iii) development-check results $r_g$. The audited rollback-integrity rate is
\[
\mathrm{RIR}=
\frac{\sum_{h\in\mathcal{H}_{\mathrm{rb}}}
\mathbb{I}[\mathrm{Restored}(h)]}
{|\mathcal{H}_{\mathrm{rb}}|}.
\]
This three-snapshot comparison does not prove restoration of the complete IR,
verification reports, process state, or append-only event history.

\section{Repair Records and Aggregate-to-Case Linkage}
\label{app:repair-locality}

The aggregate values in Table~\ref{tab:audit-metrics} are computed from
machine-readable trace and repair records. Table~\ref{tab:repair-log-schema}
identifies the fields required to audit that aggregation.

\begin{table}[!ht]
\centering
\scriptsize
\setlength{\tabcolsep}{2.5pt}
\renewcommand{\arraystretch}{1.04}
\begin{tabularx}{\columnwidth}{@{}p{0.30\columnwidth}Y@{}}
\toprule
\textbf{Field} & \textbf{Meaning} \\
\midrule
\texttt{task\_id}, record source & Identifiers joining the benchmark record,
construction trace, and source-manifest entry. \\
\texttt{failure\_type} & Symptom of the global run on $\mathcal Q$: wrong answer, exception, timeout,
memory/resource failure, or another harness verdict. \\
\texttt{evidence\_type} & Locator rule and supporting findings: registered
traceback frame, resource signal, compatible local rejection, or abstention. \\
\texttt{selected\_unit} & Responsibility node or branch root returned by
scope selection, with the locator's attribution grade $\kappa$. \\
\texttt{repair\_region}, \texttt{frozen\_region} & Editable and frozen
node IDs in the transaction's pre-repair graph. \\
\texttt{RRS\_nodes}, \texttt{CCR\_LOC} & Declared editable fraction of
pre-repair graph nodes; persisted changed-code ratio. \\
\texttt{repair\_success}, \texttt{regression\_rate} & Post-repair outcome on the
hidden APPS tests and failures among previously passing hidden tests. \\
Acceptance record and ranks & Recorded decision and the initial/candidate
strict-improvement tuples used to audit it. \\
Initial/final code, snippets, and development-check results & The three persisted snapshot
pairs used by the rollback-integrity check. \\
\bottomrule
\end{tabularx}
\caption{Per-task fields used by the reported repair metrics. The available
records support localization, persisted-delta, decision-compliance, regression,
and three-snapshot rollback checks; they do not establish a complete append-only
transaction history.}
\label{tab:repair-log-schema}
\end{table}

\begin{table}[!ht]
\centering
\scriptsize
\setlength{\tabcolsep}{2.5pt}
\begin{tabularx}{\columnwidth}{@{}p{0.31\columnwidth}p{0.08\columnwidth}Y@{}}
\toprule
\textbf{Aggregation unit} & \textbf{$n$} & \textbf{Reconciliation} \\
\midrule
Audit records & 200 & The 50/100/50
introductory/interview/competition records of the \texttt{deepseek-v4-flash}
run; no pipeline exception is reported. \\
Repair cases & 52 & Forty-four contain a recorded node/branch localization;
for the remaining eight, the locator abstains and no repair is attempted. \\
Localized repairs & 44 & Thirty pass the hidden APPS tests after repair. \\
Acceptance decisions & 44 & Thirty are accepted and 14 are rolled back. \\
Regression observations & 131 & Eligible hidden APPS tests that passed before
localized repair; none fails afterward. \\
Comparable rollbacks & 14 & All 14 restore the three audited snapshots. \\
\bottomrule
\end{tabularx}
\caption{Denominators used by the descriptive audit analysis.}
\label{tab:audit-denominators}
\end{table}

\paragraph{Scope-to-snapshot linkage.}
To reproduce RRS, the declared scope and its frozen complement are
joined to the transaction's pre-repair graph. A scope stored as a branch
root is expanded using that snapshot's hierarchy, and recorded
integration expansions are included before the ratio is computed.
Candidate-only replacement IDs are not added to either the scope count
or its denominator. These are requirements on the audit linkage, not
claims of additional serialized fields or complete-state snapshots.

\paragraph{Injection-to-scope linkage.}
The containment analysis in
Appendix~\ref{app:fault-owner-containment} uses a separate injected-fault
cohort, not the 44 natural-failure transactions summarized above.
Reproducing a containment judgment requires a task-level linkage
between the injected code location, its fault-owner ID, the exact
inspected scope, the graph/ownership snapshot, the abstention status,
and the membership judgment. The fault owner is the injection target,
not the locator's predicted responsibility. Existing fields such as
\texttt{selected\_unit} and \texttt{repair\_region} are potential sources
for the inspected scope; their correspondence to the manually checked
object must be established from the inspection record. If a scope is
stored as a branch root, its node set is interpreted in the recorded
hierarchy, not in a later replanned graph. These are logical audit
quantities, not newly asserted JSON field names or additional human
annotation results.


\section{Compact Audit Case Study}
\label{app:case-study}

This case study uses APPS task 3103 (Kattis
\texttt{helpfulcurrents}). The task counts routes from a ship's starting column
to the castle tile \texttt{@}. A ship may move north from a valid cell and may
additionally follow an east/west current on \texttt{>} or \texttt{<}. The first
assembled program returned \texttt{1} for an example whose expected output was
\texttt{2}. The trace attributes the compatible local rejections and global
wrong-answer signals to branch \texttt{S1}; the repair transaction therefore
replans that branch while retaining the outer boundary.

\begin{table}[!ht]
\centering
\scriptsize
\setlength{\tabcolsep}{2.5pt}
\renewcommand{\arraystretch}{1.04}
\begin{tabularx}{\columnwidth}{@{}p{0.29\columnwidth}YY@{}}
\toprule
\textbf{Audit question} & \textbf{Before repair} & \textbf{After repair} \\
\midrule
What failed? & Expected \texttt{2}; observed \texttt{1}. & The three task
examples pass. \\
Where is responsibility assigned? & Compatible findings map to branch
\texttt{S1}. & The transaction remains attached to \texttt{S1}; leaves remain
\texttt{S1\_S1} and \texttt{S1\_S2}. \\
What may change? & Only the implicated subtree. & The external branch
contract and frozen complement are retained. \\
What semantic change is recorded? & North/current choice and zero-based input
were under-specified or mishandled. & Parsing and row-wise DP implement the
clarified semantics. \\
\bottomrule
\end{tabularx}
\caption{Before/after audit interpretation for APPS 3103.}
\label{tab:case-study-audit-summary}
\end{table}

\begin{table}[!ht]
\centering
\scriptsize
\setlength{\tabcolsep}{1.5pt}
\renewcommand{\arraystretch}{1.03}
\begin{tabularx}{\columnwidth}{@{}P{0.42\columnwidth}P{0.16\columnwidth}P{0.23\columnwidth}Y@{}}
\toprule
\textbf{Input} & \textbf{Expected} & \textbf{Actual after repair} & \textbf{Pass} \\
\midrule
\texttt{2 2 0; >@; >\string~} & \texttt{2} & \texttt{2} & true \\
\texttt{3 5 1; >{}>@<{}<; >\string~\#\string~<; >{}>{}>{}>\string~} &
\texttt{4} & \texttt{4} & true \\
\texttt{3 4 0; >\string~@\string~; \string~<\#\string~; >{}>{}>\string~} &
\texttt{begin repairs} & \texttt{begin repairs} & true \\
\bottomrule
\end{tabularx}
\caption{Post-repair checks on the three task examples of APPS 3103.}
\label{tab:case-study-final-tests}
\end{table}

The saved trace does not preserve every internal LLM message or temporary source
file. It retains the final IR, leaf code and local verification records,
assembled code, original failing case, selected error branch, retry strategy,
and final development-check findings for the three task examples; the
planner-derived boundary cases of $\mathcal Q$ are not retained in this record. The case therefore illustrates a
bounded repair transaction, but it does not establish complete internal-state
capture or correctness beyond the retained tests.

\begin{figure*}[t]
\begingroup
\centering
\scriptsize
\setlength{\fboxsep}{4pt}
\begin{minipage}{0.94\textwidth}
\centering
\fbox{\parbox{0.70\textwidth}{\centering
\textbf{S0: outer orchestration and output contract}\\
\textit{frozen structural boundary during the selected transaction}}}

\vspace{0.2em}
$\swarrow$\hspace{0.33\textwidth}$\searrow$
\vspace{0.1em}

\begin{minipage}[t]{0.64\textwidth}
\centering
\fbox{\begin{minipage}[t]{0.94\linewidth}
\textbf{Selected bounded region $\mathcal{R}$: subtree rooted at S1}\\[-0.1em]
\textbf{S1 commitment:} parse the branch input and compute the route count while
preserving the external \texttt{raw\_input -> result} contract.\\[0.2em]
\begin{minipage}[t]{0.47\linewidth}
\textbf{S1\_S1}\quad String\\
\texttt{parse\_input}\\
records $Y,X,x_{init},grid$
\end{minipage}\hfill
\begin{minipage}[t]{0.47\linewidth}
\textbf{S1\_S2}\quad DP\\
\texttt{count\_paths}\\
row-wise route propagation
\end{minipage}
\end{minipage}}
\end{minipage}
\hfill
\begin{minipage}[t]{0.30\textwidth}
\centering
\fbox{\parbox{0.88\linewidth}{\centering
\textbf{Frozen complement $\overline{\mathcal{R}}$}\\
non-selected responsibilities are reused unchanged; any required cross-boundary
edit must be recorded as an integration exception.}}
\end{minipage}
\end{minipage}
\caption{Responsibility boundary used in the APPS 3103 audit transaction. The
selected scope is the subtree rooted at \texttt{S1}; its parser and DP leaves are
regenerated under the existing external branch contract. The outer
orchestration and non-selected responsibilities form the frozen complement.
The figure visualizes the repair transaction, not the entire internal task graph.}
\label{fig:case-study-task-graph}

\endgroup
\end{figure*}

\begin{table*}[t]
\begin{minipage}{\textwidth}
\centering
\scriptsize
\setlength{\tabcolsep}{3pt}
\renewcommand{\arraystretch}{1.03}
\begin{tabularx}{\textwidth}{@{}p{0.18\textwidth}p{0.31\textwidth}Y@{}}
\toprule
\textbf{Stage} & \textbf{Input / retained record} & \textbf{Output or audit artifact} \\
\midrule
Initial responsibility graph & Root \texttt{S0}; selected branch candidate
\texttt{S1}; parser leaf \texttt{S1\_S1}; DP leaf \texttt{S1\_S2}. &
The branch boundary is \texttt{raw\_input -> result}; the intended complexity is
$O(YX)$. \\
Local generation and checks & Strict leaf interfaces, route-counting hints, and
boundary cases. & Both leaf implementations compile, but local verification
returns \texttt{reject}; the initial plan also mishandles zero-based input and
the optional current transition. \\
Global failure & Assembled code and the problem I/O contract. &
\texttt{wrong\_answer}; for input \texttt{2 2 0\textbackslash n>@\textbackslash n>\string~},
expected \texttt{2}, observed \texttt{1}. \\
Localization & Global failure plus local verification records. &
Select \texttt{S1} using compatible local-rejection findings,
attribution grade \textsc{Low}; set \texttt{error\_branch\_root=S1}. \\
Bounded repair & Failing sample, old branch record, and frozen external boundary. &
Retain zero-based \texttt{x\_init}; make north movement available on valid cells;
treat following \texttt{>} or \texttt{<} as an additional option; regenerate only
the selected subtree and graft it back into the graph. \\
Final validation & Reassembled program and the three task examples. &
\texttt{all\_passed=true}, \texttt{cases\_run=3},
\texttt{cases\_passed=3}; the repair record is closed under the same branch ID. \\
\bottomrule
\end{tabularx}
\caption{Stage-level findings for the APPS 3103 repair transaction.
The table summarizes the retained inputs, transformations, findings, and
outcome at each stage.}
\label{tab:case-study-io}
\end{minipage}
\end{table*}

\section{Failure Modes and Applicability Boundaries}
\label{app:failure-analysis}

This appendix provides the detailed applicability analysis summarized in
Section~\ref{subsec:suitability-analysis}. The graph is most useful when
intermediate artifacts are semantically meaningful, locally testable, and
compositional. When a problem is governed by one tightly coupled global
invariant, the appropriate auditable unit may be a single global node rather
than an artificially decomposed collection of subgoals.

\begin{figure*}[t]
    \centering
    \includegraphics[width=0.97\textwidth]{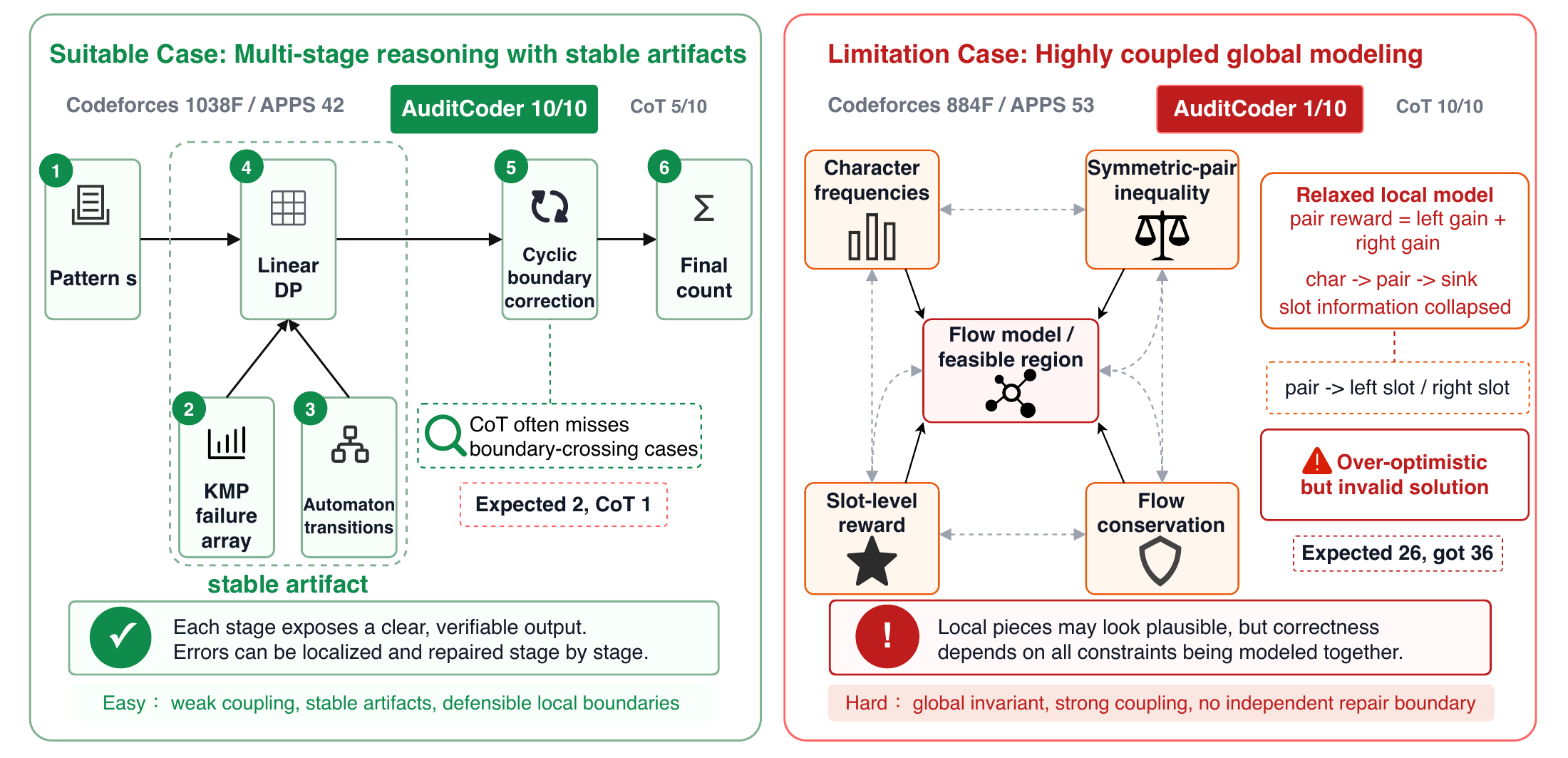}
    \caption{Two diagnostic cases illustrating the scope of graph-based responsibility boundaries. Stable, locally testable artifacts support localized validation and repair (left), whereas a tightly coupled global invariant makes locally plausible components misleading (right). The counts summarize 10 attempts for each case and are not aggregate estimates.}
    \label{fig:suitability-cases-app}
\end{figure*}


\paragraph{Diagnostic applicability cases.}
The suitable case contains a sequence of stable intermediate artifacts:
parsing, automaton construction, dynamic programming, boundary correction,
and aggregation. An implementation error can therefore remain a node-level
transaction, while an interface or decomposition error can be lifted to a
branch without changing its external contract. The limitation case contains
character frequencies, symmetric-pair constraints, slot-level rewards, and
flow conservation that jointly define one feasible region. A local surrogate
discards state needed across these constraints, so a defensible response is
to use a coarser responsibility unit, revise the graph, or abstain.

The left and right panels each summarize 10 attempts and are intended to
illustrate the structural mechanism rather than estimate the frequency of
either regime. In particular, the 10/10 versus 1/10 counts should not be
reported as aggregate task-level performance. They motivate the failure
taxonomy below and delimit when graph-based responsibility boundaries are
likely to be useful.

\begin{table*}[t]
\begin{minipage}{\textwidth}
\centering
\scriptsize
\setlength{\tabcolsep}{3.8pt}
\renewcommand{\arraystretch}{1.03}
\begin{tabularx}{\textwidth}{@{}p{0.17\textwidth}p{0.37\textwidth}Y@{}}
\toprule
\textbf{Failure mode} & \textbf{Observed signal} & \textbf{Response} \\
\midrule
Local implementation bug &
A registered traceback frame or compatible local contract/boundary rejection
points to one implementation owner. &
Prefer node-level repair; freeze the interface and the outside region unless
findings show a contract mismatch. \\
Contract or interface mismatch &
Adjacent nodes disagree on arguments, return schema, shape, or boundary
conditions. &
Repair the selected contract/code unit and expand the transaction only as needed
to preserve the external frozen boundary. \\
Incorrect decomposition &
Local goals appear individually plausible but omit a dependency or assign the
wrong responsibility. &
Lift to branch-level repair and replan within the branch's external input/output
contract. \\
Global invariant or high coupling &
Correctness depends on constraints that must be modeled jointly, and no stable
local boundary is justified. &
Revise the graph abstraction or use one global responsibility node; bounded
repair may abstain. \\
Ambiguous localization &
Several incomparable nodes remain compatible with the findings, or the findings
establish only a graph-level violation. &
Return $\bot$ rather than assigning an unsupported unit. \\
Repair-induced regression &
A previously passing test fails after the transaction. &
Record the non-monotonic outcome; accept the candidate only if its
development-check rank strictly improves; inspect boundary compatibility and
integration exceptions. \\
\bottomrule
\end{tabularx}
\caption{Failure taxonomy and bounded repair response.}
\label{tab:failure-taxonomy}
\end{minipage}
\end{table*}

\section{Boundary Constraints and Artifact Map}
\label{app:boundary-prompt-map}

Table~\ref{tab:boundary-prompt-requirements} connects the lifecycle
requirements to the components that implement them. The procedures
are specified in Algorithms~\ref{alg:auditcoder}--\ref{alg:coherent};
this section summarizes their boundary and artifact requirements
without repeating prompt templates.

\begin{table}[!ht]
\centering
\scriptsize
\setlength{\tabcolsep}{2.5pt}
\renewcommand{\arraystretch}{1.02}
\begin{tabularx}{\columnwidth}{@{}p{0.19\columnwidth}Y@{}}
\toprule
\multicolumn{2}{@{}l@{}}{\textbf{Task decomposer}} \\
\textbf{Requirement} & Allocate stable node IDs; declare parent/dependency
relations, input/output slots, provenance sources, leaf status, and complexity
budgets before code generation. \\
\textbf{Audit role} & Creates the pre-code responsibility address space. \\[1pt]
\multicolumn{2}{@{}l@{}}{\textbf{Interface designer}} \\
\textbf{Requirement} & Produce a strict function signature, typed parameters,
return schema, preconditions, and postconditions for every leaf code owner. \\
\textbf{Audit role} & Defines the machine-addressable contract attached to the
responsibility ID. \\[1pt]
\multicolumn{2}{@{}l@{}}{\textbf{Node-local generator}} \\
\textbf{Requirement} & Implement only the assigned interface and return one
node-owned function/helper bundle. \\
\textbf{Audit role} & Prevents a generation call from silently acquiring another
node's responsibility. \\[1pt]
\multicolumn{2}{@{}l@{}}{\textbf{Assembler and ownership checker}} \\
\textbf{Requirement} & Assemble in dependency order, pass values using provenance
links, preserve child snippets, and convert any necessary child edit into an
integration exception. \\
\textbf{Audit role} & Maintains the total function-to-owner map and makes
cross-boundary edits auditable. \\[1pt]
\multicolumn{2}{@{}l@{}}{\textbf{Failure locator}} \\
\textbf{Requirement} & Apply registered traceback, resource signals, compatible
local rejection, and abstention in fixed priority order. \\
\textbf{Audit role} & Identifies implicated responsibilities; scope selection
uses these findings to choose an editable node or branch. \\[1pt]
\multicolumn{2}{@{}l@{}}{\textbf{Branch replanner}} \\
\textbf{Requirement} & Replan only the selected branch; preserve its external
input, output, return-type, and downstream-call compatibility; do not replan the
frozen complement. \\
\textbf{Audit role} & Makes the selected repair boundary executable rather than
advisory. \\[1pt]
\multicolumn{2}{@{}l@{}}{\textbf{History writer}} \\
\textbf{Requirement} & Store findings, attribution grade, selected/frozen regions,
graph/code deltas, integration exceptions, and post-repair result. \\
\textbf{Audit role} & Appends the repair transaction to the same responsibility
record. \\
\bottomrule
\end{tabularx}
\caption{Boundary-critical component requirements and their audit roles.
The table summarizes the method specification; it is not an additional
measurement of implementation compliance.}
\label{tab:boundary-prompt-requirements}
\end{table}

\paragraph{Boundary enforcement.}
Replanning is subject to the final declared scope and its exposed
boundary contracts (Appendix~\ref{app:boundary}).
Algorithm~\ref{alg:auditcoder} rejects undeclared cross-boundary edits.
The \emph{intended edits} and \emph{integration exceptions} in
Algorithm~\ref{alg:scope} are pre-regeneration declarations, not a
scope reconstructed retrospectively from the candidate's diff. These
constraints are part of the repair specification rather than permissions
granted by a generic replanning instruction.


\paragraph{Artifact-to-result linkage.}
The reproducibility specification connects each reported result to
its fixed task set, configuration, generated artifacts, evaluation
script, and retained records. Source and prompt versions identify the
construction procedure; ownership and repair records support the
case-level checks described in Appendix~\ref{app:repair-locality}.
An aggregate table is not a substitute for those records.


\section{Exploratory Cross-Domain Evaluation Suites}
\label{app:cross-domain-suites}

\paragraph{Purpose and scope.}
We use three small executable suites to check whether the generation
pipeline can operate beyond competitive-programming and class-level
Python tasks while retaining structured audit records. The 22 tasks span
file-oriented financial analytics, scientific numerical computing, and
aerospace/communications primitives. The suites are deliberately
heterogeneous and are selected or authored rather than random samples
of their domains. Their results are therefore descriptive: one task
changes a suite-level rate by 12.5--16.7 percentage points, so
resolved-task counts---not percentage differences---should guide
interpretation. The study does not estimate broad domain-level
generalization or establish a statistically reliable ranking between
methods.

\paragraph{Protocol.}
QF-Small6 requires complete programs that read public data files and
produce structured output files; SciCode Blind-Small-8 and
AeroComm-Small8 require importable Python functions. For each task,
AgentCoder, \method{}, MapCoder, and CodeSim receive the same public
specification and visible input data, use the same
\texttt{deepseek-v4-flash} base model, return
one final candidate, and receive no feedback from the frozen hidden
tests. A task is resolved only when the candidate imports or exits
successfully and passes every test for that task. Test-level pass counts
are retained for diagnosis but are not pooled across suites because the
number and granularity of tests differ.

\begin{table*}[!t]
\caption{Composition of the exploratory cross-domain suites. Code scale
refers to the upstream reference implementation for QF-Small6 and the
expected implementation range for AeroComm-Small8; SciCode does not
expose reference implementations.}
\label{tab:cross-domain-datasets}
\centering
\small
\setlength{\tabcolsep}{4pt}
\begin{tabularx}{\textwidth}{@{}lccclX@{}}
\toprule
Suite & Tasks & Unit & Frozen tests & Code scale & Domain coverage \\
\midrule
QF-Small6
& 6 & File program & 64
& 42--62 LOC
& Earnings analysis, corporate-action adjustment, historical VaR,
interest-rate derivatives, yield-curve bootstrapping, and portfolio
factors \\
SciCode Blind-Small-8
& 8 & Function & 24
& Not exposed
& Electromagnetics, biophysics, ecology, optics, condensed matter,
thermal engineering, and molecular dynamics \\
AeroComm-Small8
& 8 & Function & 72
& 6--22 LOC
& Orbital mechanics, spacecraft attitude, radio propagation,
modulation, error detection, and channel coding \\
\bottomrule
\end{tabularx}
\end{table*}

\begin{figure}[!t]
    \centering
    \includegraphics[width=\columnwidth]
    {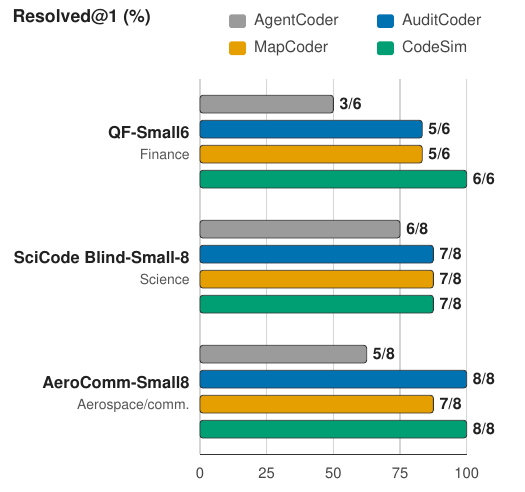}
    \caption{Exploratory cross-domain fixed-run
    \textsc{Resolved@1} under frozen hidden tests using
    \texttt{deepseek-v4-flash}, comparing AgentCoder, \method{},
    MapCoder, and CodeSim. A task is resolved only when the final
    retained candidate passes every test. Labels report resolved tasks
    over suite size; because each suite contains only six or eight tasks,
    the counts rather than percentage differences should guide
    interpretation.}
    \label{fig:cross-domain-results}
\end{figure}

\paragraph{Functional results.}
Under this protocol, AgentCoder, \method{}, MapCoder, and CodeSim
resolve 3/6, 5/6, 5/6, and 6/6 QF-Small6 tasks; 6/8, 7/8, 7/8, and
7/8 SciCode Blind-Small-8 tasks; and 5/8, 8/8, 7/8, and 8/8
AeroComm-Small8 tasks, respectively (Figure~\ref{fig:cross-domain-results}).
\method{} exceeds AgentCoder on all three suites, ties MapCoder on
finance and scientific computing, and resolves one more
aerospace/communications task than MapCoder. It also ties CodeSim on
SciCode and AeroComm, while resolving one fewer finance task.
The results extend \method{}'s functional competitiveness to file-based
and numerical Python programs under heterogeneous execution forms.
Given the small denominators, these outcomes support exploratory
cross-domain applicability rather than a statistically reliable
domain-level ranking.

\paragraph{Audit-record checks.}
For QF-Small6, SciCode Blind-Small-8, and AeroComm-Small8,
respectively, \method records task-macro DCTC scores of 0.893, 0.912,
and 0.935. Across all 22 records, every task graph passes the
structural-integrity check, and 33 of 35 code-owning nodes satisfy
contract--signature consistency. DCTC measures responsibility-record
completeness rather than semantic correctness and complements the
functional resolution counts in Figure~\ref{fig:cross-domain-results}.
These checks concern final trace structure; the cross-domain study
does not evaluate localization, bounded-repair success, or
frozen-complement preservation for these suites.

\paragraph{QF-Small6: file-oriented finance tasks.}
QF-Small6 is a six-task subset of QuantitativeFinance-Bench
\citep{quantitativefinancebench2026}, frozen from upstream revision
\texttt{024921e} under the CC BY-NC 4.0 license. The subset combines
short implementations with realistic financial data and deterministic
domain invariants. It covers earnings-surprise and
standardized-unexpected-earnings calculation, split/dividend adjustment,
historical VaR data preparation, Black-style cap/floor pricing,
zero-coupon curve bootstrapping, and PCA-based factor-neutral portfolio
construction. Tasks read public CSV or JSON inputs and produce JSON or
CSV artifacts; the upstream reference implementations contain
approximately 42--62 nonblank, noncomment lines.

The evaluator combines 48 upstream tests with 16 additional contract
cases covering required output files, data cleaning, financial
identities, numerical tolerances, and output schemas. A candidate must
terminate successfully, produce every required artifact, and pass both
groups. The evaluated package is versioned as
\texttt{qf-small6-adag-v1.2}, where \texttt{adag} is the internal name of
the \method{} implementation. It adds four public clarifications for
observable ambiguities: the top-level wrappers of the earnings and
corporate-action JSON files, the adjacent-maturity convention for
forward rates on a non-contiguous maturity grid, and the representation
of target weights and sample-standard-deviation convention in the PCA
task. Original upstream statements are retained separately, while
private tests and oracle implementations remain inaccessible to the
generator.

\paragraph{SciCode Blind-Small-8: blind scientific subproblems.}
SciCode Blind-Small-8 contains eight independently evaluable subproblems
from the SciCode test split~\citep{tian2024scicode}, frozen at dataset
revision \texttt{4510f6a} and evaluator revision \texttt{e3158ea}. Each
task is the first independent substep of its parent problem, exposes no
reference implementation, and requires one Python function. The public
prompt includes the scientist-provided substep description and
background needed to make the task self-contained. The suite covers a
Maxwell-equation finite-difference operator (\texttt{13.1}),
optical-tweezer Brownian dynamics (\texttt{14.1}), chemostat species
growth (\texttt{25.1}), Gaussian-beam propagation (\texttt{28.1}),
quantum-dot absorption wavelength (\texttt{35.1}), paraxial ray
propagation through a compound lens (\texttt{37.1}), heat-equation grid
initialization (\texttt{45.1}), and periodic-boundary coordinate
wrapping (\texttt{77.1}).

Each task uses three official numerical cases, for 24 frozen tests in
total. Test inputs, numerical targets, and the target HDF5 subset are
excluded from the generation prompt. Before any method was run, a
specification--target audit removed two candidate tasks:
\texttt{24.1} used an inconsistent branch at $x=0$, and \texttt{76.1}
used an L2-normalized target despite specifying L1 normalization. They
were replaced by \texttt{13.1} and \texttt{77.1} under the same
selection criteria, and the final task list was hashed before model
evaluation.

\paragraph{AeroComm-Small8: source-grounded engineering primitives.}
AeroComm-Small8 is an authored, source-grounded suite rather than an
official NASA, ITU, or CCSDS benchmark. It contains four aerospace and
four communications tasks, each requiring one importable Python
function using only the Python standard library and NumPy. The aerospace
group covers vis-viva orbital speed, coplanar Hohmann-transfer
budgeting, numerical solution of the elliptic Kepler equation, and
quaternion-based spacecraft attitude rotation. The communications group
covers Gray-coded QPSK mapping, free-space path loss, CCSDS-style
CRC-16 calculation, and rate-$1/2$ convolutional encoding with octal
generators $(7,5)$.

The specifications were written from public formulas, standards, and
engineering primitives documented by NASA/JPL, ITU-R, CCSDS, CommPy,
and Komm. A frozen source registry records which sources support each
task; the suite should therefore be described as source-grounded rather
than as an official subset of those organizations' software. Expected
implementations range from approximately 6 to 22 nonblank lines. The
evaluator contains 72 hidden cases covering nominal values, boundary
conditions, vectorization, dtype and shape requirements, invalid inputs,
and numerical tolerances. All eight private oracle implementations pass,
whereas a deterministic negative control resolves none of the tasks.
The suite evaluates short numerical and coding primitives; it does not
represent safety-critical flight software, hard real-time execution,
hardware-in-the-loop validation, or a complete communications protocol
stack.